\newcommand{\fix}{\mathrm{fix}}

\documentclass[times, twoside]{zHenriquesLab-StyleBioRxiv}

\begin{document}

\leadauthor{Koldaeva} 

\title{Population genetics in microchannels}

\author[a]{Anzhelika Koldaeva}
\author[b]{Hsieh-Fu Tsai}
\author[b]{Amy Q. Shen}
\author[a,\Letter]{Simone Pigolotti}

\affil[a]{Biological Complexity Unit, Okinawa Institute for Science and Technology and Graduate University, Onna, Okinawa 904-0495, Japan}

\affil[b]{Micro/Bio/Nanofluidics Unit, Okinawa Institute for Science and Technology and Graduate University, Onna, Okinawa 904-0495, Japan}

\maketitle

\begin{abstract}
Spatial constraints such as rigid barriers affect the dynamics of cell populations, potentially altering the course of natural evolution. In this paper, we investigate the population genetics of \textit{Escherichia coli} proliferating in microchannels with open ends. Our analysis is based on a population model in which reproducing cells shift entire lanes of cells towards the open ends of the channel. The model predicts that diversity is lost very rapidly within lanes, but at a much slower pace among lanes. As a consequence, two mixed, neutral \textit{E.~coli} strains competing in a microchannel must organize into an ordered regular stripe pattern in the course of a few generations. These predictions are in quantitative agreement with our experiments. We also demonstrate that random mutations appearing in the middle of the channel are much more likely to reach fixation than those occurring elsewhere. Our results illustrate fundamental mechanisms of microbial evolution in spatially confined space. 
\end {abstract}


\begin{corrauthor}
simone.pigolotti@oist.jp
\end{corrauthor}

\section*{Introduction}
Biological populations can be spatially organized by landscape barriers that constrain individual movement, generating ordered patterns at the population level.  For example, populations of rod-shaped bacteria \textit{Escherichia coli} growing on surfaces tend to organize into domains of aligned cells \cite{geometry_and_mechanic,dell2018growing}. When \textit{E. coli} grows in confined channels, cell alignment is affected by the geometry of the channel boundaries \cite{self_organization,karamched2019moran}. In narrow channels, populations reach a highly ordered structure, with cells organized parallel to each other and to the boundaries \cite{nemat_align, bomechanical_ordering}. In wide channels, such alignment is disrupted at large scales by a buckling instability \cite{buckling}. 

Once a cell population densely fills a microchannel, dividing cells push others toward the open ends, potentially leading to expulsion of cells. The timescale at which cells are expelled is typically shorter than their lifetime  (for an estimate of the latter see, e.g., \cite{rob_Ecoli}). Therefore, death events can be usually neglected when focusing on a population inside a microchannel.  In genetically diverse populations, we expect such competition to reduce diversity at a pace that depends on the channel dimensions relative to the cell size.

Microfluidic devices constitute ideal experimental systems to study population growth in confined geometries \cite{levien2020non}. Size and shape of the microchannels in these devices can be tailored to mimick microorganism habitats ~\cite{micrfl_exp_front, micrfl_env}. Nutrients can be delivered to the residing microorganisms inside the microchannels by controlled flows. Such devices are often used to track population dynamics of microorganisms at the single-cell level over several generations \cite{Bacterial_growth, Streaming_Instability}. A paradigmatic example is the ``mother machine'' -- a microchannel with one open end and small enough width to accommodate a single lane of cells \cite{rob_Ecoli}. Microchannels with two open ends have been used to validate a relation between the cell division time distribution and the population growth rate (\cite{hashimoto}, but see \cite{lin2017effects}).

From the theoretical side, competition in confined geometries has been scarcely studied. Common spatial competition models, ranging from Kimura's stepping stone model \cite{Kimura_Weiss, step_stone} to generalizations in the theory of evolutionary graphs \cite{evolutionary_dynamics_on_graphs, spat_pop_rew,nowak_evolutionary_dynamics,voter_m_on_heter}, include birth, death, and diffusion events only, and do not account for cell-to-cell mechanical interactions in confined geometries.  Single-lane models in which newborn individuals shift their neighbors away have been theoretically studied in the context of cancer progression \cite{Iwasa,allen2012evolutionary}. Microbial population dynamics model incorporating shifting has been investigated with computational approaches \cite{hashimoto, line_form}. In particular, simulations of a model of competing bacterial strains \cite{line_form} show a formation of lanes along the channel axial direction. Theoretical predictions for geometries hosting multiple lanes and quantitative experimental validations have been lacking.

In this paper, we study the population genetics of bacterial colonies growing in confined geometries. We combine theory, numerical simulations, and experiments on \textit{E.~coli} populations growing in rectangular microchannels with two open ends. We introduce our work by first presenting experiments in which two fluorescently-marked neutral {\it E. coli} strains competing in microchannels demix into a stripe pattern. We explain this phenomenon by means of an individual-based population model describing competition between strains inside the channel. This model reveals that the diversity loss within each lane is much faster than predicted by traditional spatial population genetics, in quantitative agreement with our experimental observations.  We discuss the consequences of our results and their implications for the evolution of cell populations in confined geometries.
\section*{Results}

\begin{figure*}[htb]
\centering
\includegraphics[width=.99\linewidth]{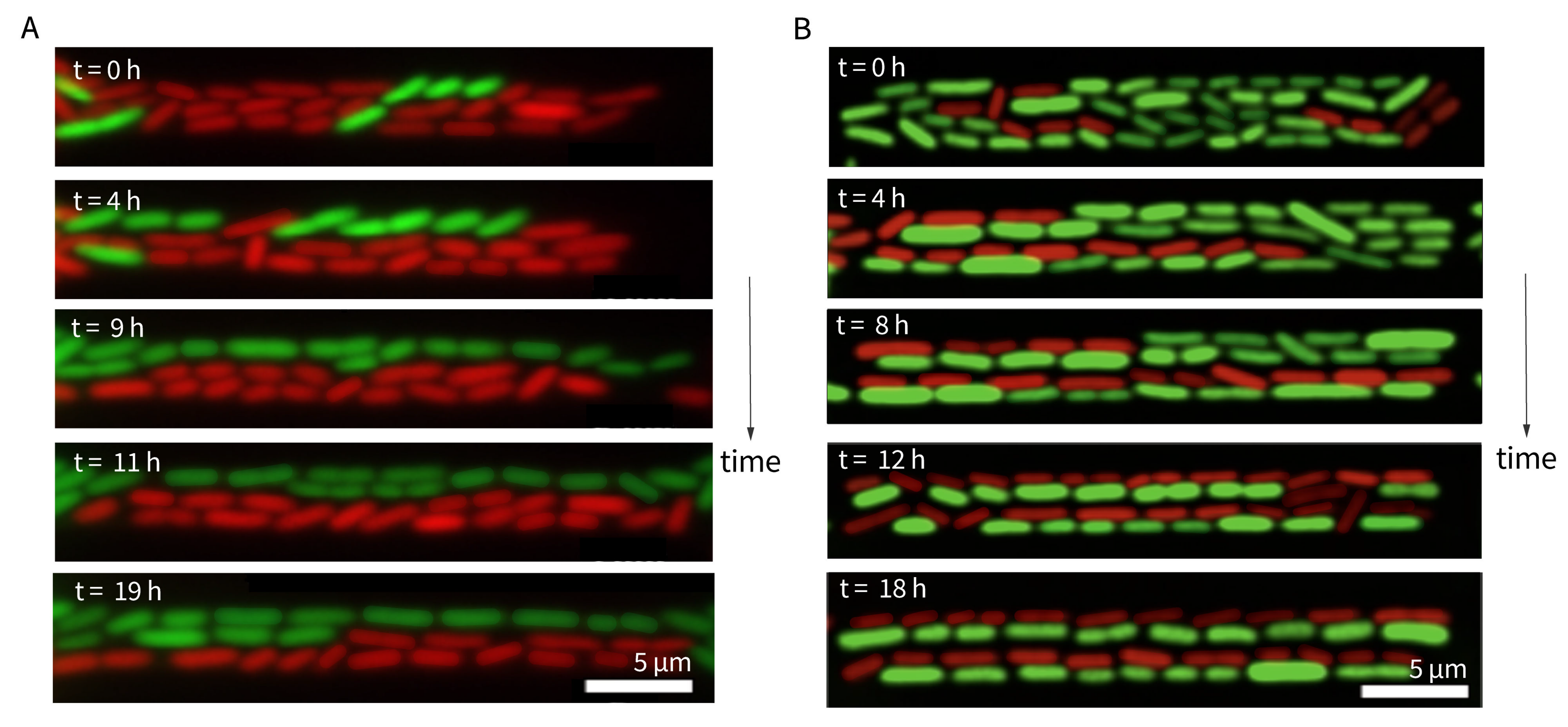}
\caption{Competition between two \textit{E. coli} strains (in red and green) in microchannels with two open ends. The two panels show two experimental realizations in microchannels of different widths. \textit{(A)} Competing strains form two stripes in a channel of width 2.5~$\mu$m harboring three lanes of cells. \textit{(B)} Strains segregate into four stripes in a channel of width 3~$\mu$m harboring four lanes of cells. The observed number of stripes fluctuates among different experimental runs, see Fig.~S1A in SI Appendix. }
\label{fig:simplif}
\end{figure*}

\subsection*{\label{subsec:Ecoli_in_micr}Competing neutral \textit{E. coli} strains form stripe patterns in microchannels}

As a motivation, we present an experiment that anticipates a main consequence of our theory. We inoculated a mixture of two \textit{E. coli} strains  into microchannels, see Fig.~\ref{fig:simplif}A and \ref{fig:simplif}B. The two strains are labeled with green and red fluorescent proteins and are otherwise neutral, i.e., have the same fitness. Bacteria reproduce in nutrient-rich conditions inside the microchannels and push each other toward the open ends. As a result, cells are continuously expelled while the number of cells inside the channels remains nearly constant. Our microchannels are rectangular, 30 $\mu$m long, 1 $\mu $m deep, and have variable width from  1 $\mu $m to  3 $\mu $m, unless specified otherwise. For comparison, \textit{E. coli} cells are $2.1 \pm 0.2$ $\mu$m in length and $0.65 \pm 0.04$ $\mu$m in width, so that the microchannels host monolayers of cells of width ranging from one to four lanes. 

In about eight hours, the two strains organize themselves into a regular stripe patterns, see Fig.~\ref{fig:simplif}A, ~\ref{fig:simplif}B, and supplementary movies 1 and 2.  The number of stripes and their width depend on the microchannel width and also fluctuates depending on the initial arrangement of inoculated cells. In these experiments, the average cell division time is of approximately 95~min, meaning that lanes are formed within a few generations. Our focus is on microchannels hosting monolayers of cells, but we also observe stripe formation in deeper (3~$\mu$m) microchannels, harboring multiple layers of cells (see supplementary movie 3).

\subsection*{\label{sec:model_symm_rep}Population model predicts the stripe pattern}

\begin{figure}[ht!]
\centering
\includegraphics[width=.99\linewidth]{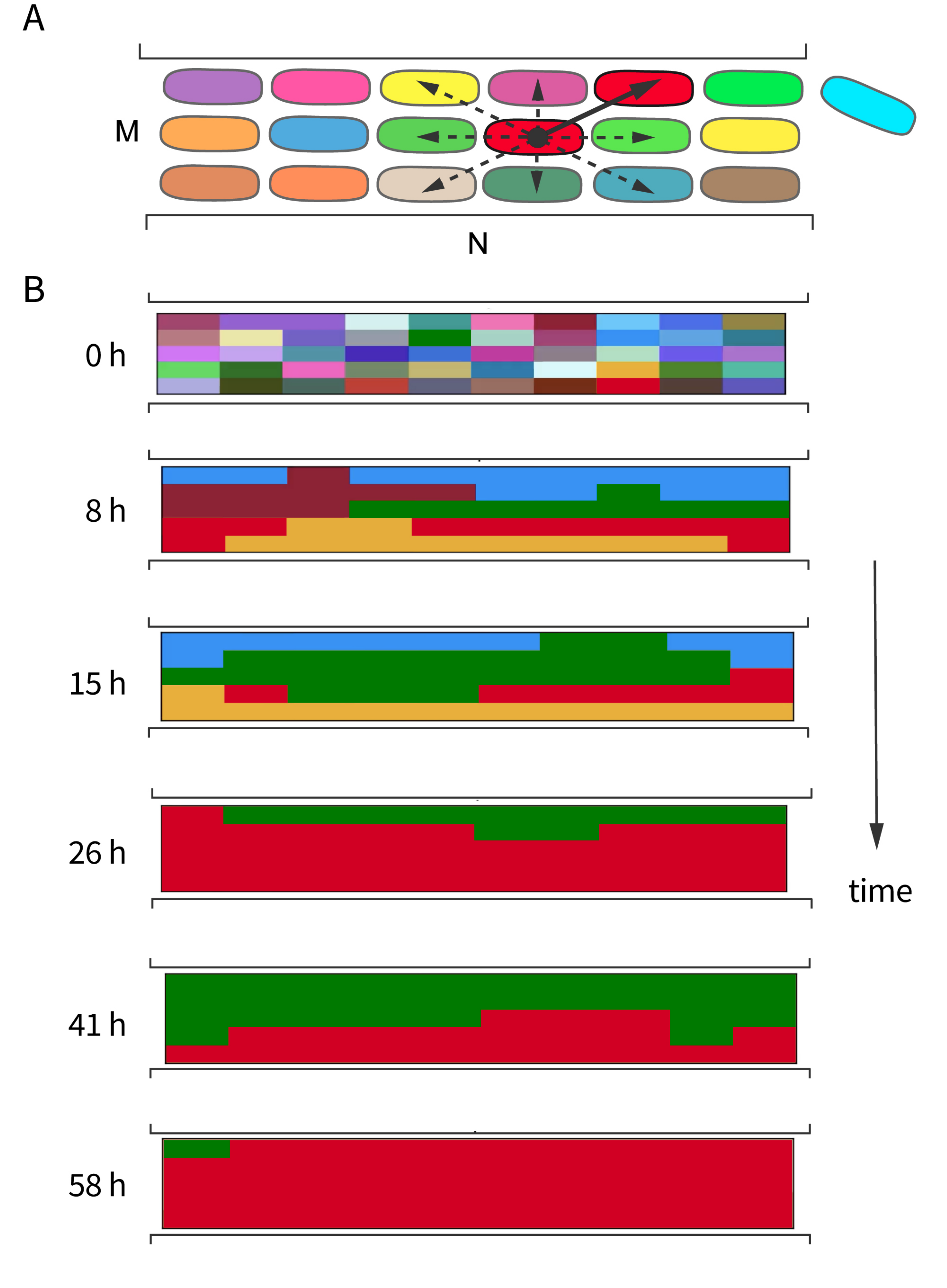}
\caption{Population model describes cells proliferating in a microchannel. \textit{(A)} Scheme of the model. Different colors represent different clonal populations. A randomly chosen cell reproduces to the right (arrow) and shifts all the cells to its right toward the right end of the microchannel. As a result, the cell next to the right end is expelled from the microchannel. Dashed arrows show the other 7 possible directions for reproduction. Cells located at the boundaries can reproduce in 5 possible directions.   \textit{(B)} Dynamics of the model. The dynamics progresses until one clonal population takes over the entire population. See Fig.~S1 in SI Appendix for a more extensive comparison between the patterns observed in experiments and in simulations. Parameters are $M=5$, $N=10$, $b = 0.01$ min$^{-1}$, $m = 0.6$ and $\alpha = 3.2$. 
}
\label{fig:simplif_mod}
\end{figure}

We want to understand how the genetic diversity of a microbial population in a microchannel changes with time. We model the microchannel as a lattice of $M \times N$ sites. Each site is always occupied by one cell, see Fig.~\ref{fig:simplif_mod}A. We define clonal populations as groups of cells that originate from a common ancestor in the initial population. The dynamics of the clonal populations permits us to determine the patterns that the population would develop if some of the cells were fluorescently marked, or carried a neutral mutation. 

The dynamics proceeds as follows. Cells reproduce binarily at a constant rate $b$. After reproduction, one daughter cell takes the position of its mother. The other occupies one of the adjacent lattice sites and shifts a lane of existing cells toward one open end of the microchannel. As a result, a cell at the open end is expelled from the microchannel. If a reproducing cell is located next to an open end, its daughter can end up outside the microchannel, thereby being immediately expelled.

Our experiments reveal that the probabilities of choosing neighboring sites are not uniform. Specifically, we identify two effects that bias these probabilities. The first effect is related to the mass of the lane of cells to be shifted. We observe that reproduction events that shift shorter lanes of cells are more likely. We quantify this effect via a mass parameter $m\ge0$. Increasing $m$ biases reproduction in the direction closer to an open end. The second effect is the preference of cells to reproduce within their same lane, due to their aligned arrangement and the rod shape of {\em E. coli}. We introduce an alignment parameter $\alpha >0$ equal to the relative probability of a reproduction event within a lane over that of an event involving a change of lane. In the limiting case $\alpha = 1$ and $m = 0$, the  reproduction probabilities are uniform as in the model numerically studied in \cite{hashimoto}.

We determine the parameters $N$, $M$, $b$, $m$, and $\alpha$ from our experiments with a single \textit{E. coli} strain in channels of different width, see Materials and Methods and SI Appendix. We find that $N$ and $b$ are the only parameters that significantly depend on the channel width, see Table~\ref{table2}.

We take the number $A(t)$ of clonal populations in the microchannel at time $t$ as our measure of diversity. At the initial time $t = 0$, we have $A(0)=MN$. Diversity decreases with time, as progenies of initial individuals are expelled from the microchannel, see Fig.~\ref{fig:simplif_mod}B. At intermediate times, the surviving strains tend to form stripe patterns that resemble those in Fig.~1  (see Fig.~S1 in SI Appendix for a more extensive qualitative comparison). The model further predicts that competition between stripes should lead to fixation of one of the strains at very long times.

\vspace{600px}

\begin{table}[tbhp]
\centering
\caption{Parameters evaluated from the experimental recordings}
\begin{tabular}{cccccc}
 width & $M$ & $N$ & $b$ (1/min) & $m$ & $\alpha$\\
 \hline
1 $\mu$m& 1 & 13 & 0.007 & 0.6 & - \\
1.5 $\mu$m& 2 & 9 & 0.007 & 0.6 & 3.2 \\
2.5 $\mu$m& 3 & 9 & 0.01 & 0.6 & 3.2\\
3 $\mu$m& 4 & 9 & 0.0105 & 0.6 & 3.2\\
 \hline
\end{tabular}
\label{table2}
\end{table}

\begin{figure*}[ht!]
\centering
\includegraphics[width=.99\linewidth]{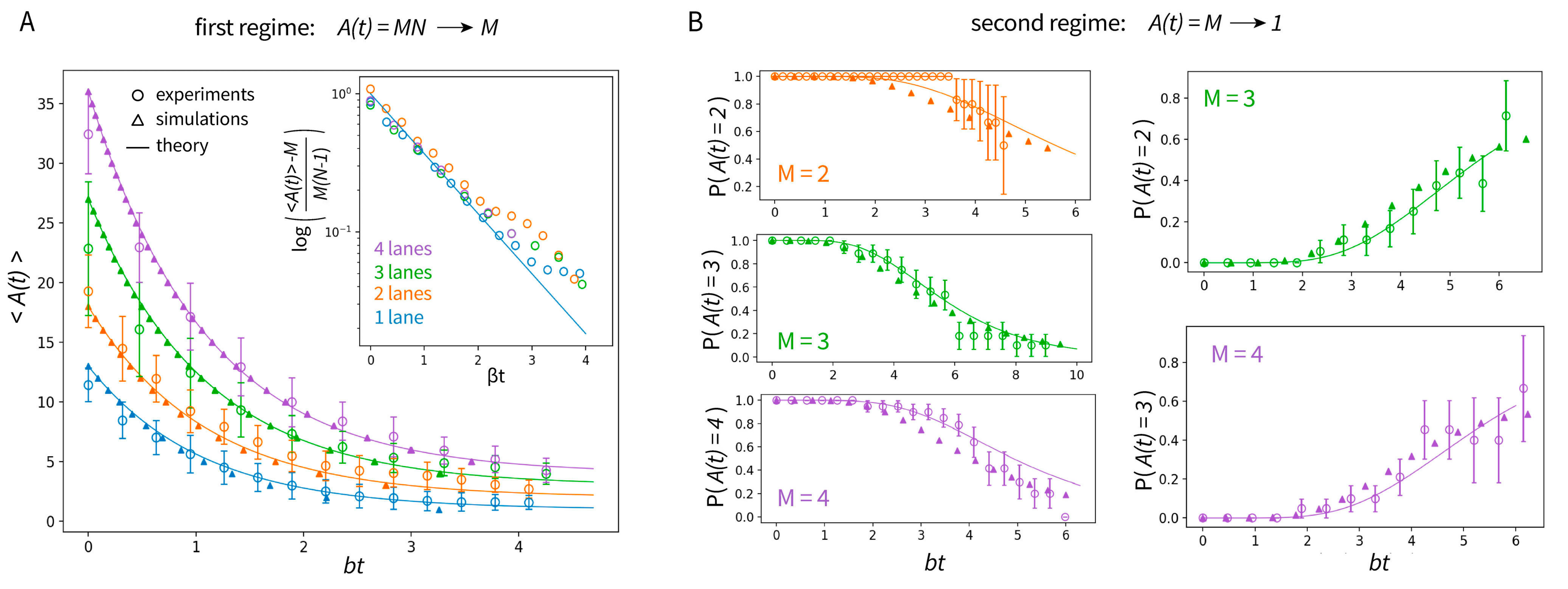}
\caption{Two temporal regimes of diversity loss. In all plots, circles with error bars represent the experimental data; triangles represent numerical simulations; and solid curves represent analytical solutions. Model parameters are listed in Table~\ref{table2}. \textit{(A)} First regime of diversity loss. Theory, simulations, and experiments show that diversity decreases exponentially in time. Time is measured in generations. The solid curves represent the analytical solutions given by \eqref{A_t_mult}.  The inset shows a linear data collapse of the experimental data based on \eqref{A_t_mult}. \textit{(B)} Second regime of diversity loss. Probability of observing a given number of clonal population as a function of time, measured in generations from the start of the second regime. The experimental data is obtained by re-tracking our experimental data, see Methods. Details on the data analysis and analytical solutions are in SI Appendix.}
\label{fig:numb_clon_popul}
\end{figure*}

\subsection*{Diversity loss and fixation is exponentially fast in single-lane microchannels}

We study diversity loss starting from microchannels with a single lane. In this case, we compute the rate of diversity loss by focusing on the interfaces between clonal populations (see Materials and Methods). We find that the average diversity at time $t$ is equal to
\begin{equation}
\langle A(t)\rangle = (N-1)e^{-\beta t}+1,
\label{A_t_one}
\end{equation}
where we define 
\begin{equation}
\beta=b\left(1-\frac{m}{N-1}\right).
\label{beta}
\end{equation}
Equation~(\ref{A_t_one}) shows that diversity loss in single-lane channels is exponentially fast. This result is in stark contrast with classic spatial population models such as the voter model, where diversity decays as $t^{-1/2}$ in one dimension \cite{bramson1980asymptotics}. The characteristic rate $\beta$ at which diversity is lost is on the order of the reproduction rate $b$, apart from a correction term that depends on the mass parameter $m$.

At long times, one clonal population eventually takes over the entire microchannel. The time at which this event occurs is called the fixation time. In microchannels with a single lane, the average fixation time is equal to 
\begin{equation}
\langle T_{N \to 1}\rangle  = \beta^{-1} \sum_{A = 2}^N \frac{1}{A-1} \approx \beta^{-1}[\log(N-1)+\gamma],
\label{fix_time_m}
\end{equation}
where $\gamma \approx 0.577$ is the Euler–Mascheroni constant, and the approximation is valid for large $N$, see Methods. The logarithmic dependence of the fixation time on the population size $N$ reflects the fact that the number of clonal populations decays exponentially in time, see \eqref{A_t_one}. The theoretical predictions of \eqref{A_t_one} and \eqref{fix_time_m} are in excellent agreement with our experiments, see Fig.~\ref{fig:numb_clon_popul}A and Table \ref{table}.  

\subsection*{First regime of diversity loss: exponentially fast fixation within each lane}

Our results in the single lane case suggest that, in microchannels with multiple lanes, competition within each lane should lead to an exponentially fast diversity loss.  In contrast, we expect competition among lanes to be less effective at reducing diversity. The alignment of cells favors reproduction events within each lane, further enhancing this difference. Following this idea, we identify two temporal regimes of diversity loss. In the first regime, diversity rapidly decreases from $A=MN$ to $A=M$, primarily due to competition within lanes. The second regime ranges from $A=M$ down to $A=1$ and is characterized by competition among lanes.

The first regime is characterized by negligible interaction among lanes. It follows from \eqref{A_t_one} that the average number of clonal populations at time $t$ is approximated by
\begin{equation}
\langle A(t)\rangle \approx M(N-1)e^{-\beta t}+M.
\label{A_t_mult}
\end{equation}
We test this prediction in experiments with a single {\it E. coli} strain, where we track descendant of each individuals in the initial population (see Materials and Methods). We find an excellent agreement, see Fig.~\ref{fig:numb_clon_popul}A. \eqref{A_t_mult} also implies that the quantity $\log[(\langle A(t) \rangle-M)/(M(N-1))]$ must be a universal linear function of $\beta t$, see inset of Fig.~\ref{fig:numb_clon_popul}A.

We approximate the average duration $\langle T_{MN \to M}\rangle$ of the first regime as
\begin{equation}
 \langle T_{MN \to M}\rangle \approx\beta^{-1}[\log(M(N-1))+ \gamma],
 \label{T_MN_N}
\end{equation}
see SI Appendix. This approximation and numerical simulations of the model agree well with our experiments, see Table \ref{table}.

\subsection*{Second regime of diversity loss: slow competition among lanes}

In the second temporal regime of diversity loss, competition among lanes becomes relevant. This competition is driven by events in which a cell reproduces in a neighboring lane and its progeny eventually colonizes the entire lane. These events occur at a rate that we estimate to be quite small, see SI Appendix. Aside from these events, lanes are typically dominated by a single clonal population, see SI Appendix. 

Following these ideas, we can consider lanes as single units which invade each other at a certain rate. This process is called invasion process in the literature \cite{voter_m_on_heter}. We mathematically solve this invasion process and thereby estimate the probability to observe a given diversity $A(t)$ in the second regime, see SI Appendix. Our experimental results agree very well with simulations of the model and qualitatively agree with the results from the invasion process, see  Fig.~\ref{fig:numb_clon_popul}B and  ~\ref{fig:numb_clon_popul}C. Our theoretical and numerical results predict that the average fixation time for microchannels with multiple lanes is very long, and therefore inaccessible in our experiments, see Table \ref{table}.

\begin{table*}[tbhp]
\centering
\caption{Mean duration of the first regime $T_{MN \to M}$ and the second regime $T_{M \to 1}$. Time in the experimental data is scaled by the division rate $b$ evaluated for each group of data with $1, 2, 3$ and $4$ lanes. The associated uncertainties are standard deviations. Parameters for the theoretical and numerical predictions are summarized in Table~\ref{table2}.}
\begin{tabular}{ccccccc}
&\multicolumn{1}{c}{}&\multicolumn{3}{c}{$T_{MN \to M}$}&\multicolumn{2}{c}{$T_{M \to 1}$}\\
microchannel width&number of lanes&experim.&theoret.&numeric.&theoret.&numeric.\\ 
 \hline
1 $\mu$m&$M = 1$&$3.58 \pm 1.4$&$3.22 \pm 1.31$&$3.26 \pm 1.21$&$-$&$-$    \\
1.5 $\mu$m&$M = 2$&$4.04 \pm 1.19$&$3.62 \pm 1.74$&$3.95 \pm 1.46$&$5.95 \pm 2.61$&$6.39 \pm 3.99$    \\
2.5 $\mu$m&$M = 3$&$4.75 \pm 0.79$&$4.05 \pm 1.83$&$4.28 \pm 1.36$&$14.68 \pm 5.56$&$13.47 \pm 8.44$    \\
3 $\mu$m&$M = 4$&$3.86 \pm 0.42$&$4.37 \pm 1.89$&$4.42 \pm 1.25$&$21.25 \pm 6.23$&$22.32 \pm 14.04$   \\
 \hline
\label{table}
\end{tabular}
\end{table*}

\subsection*{\label{sec:moth_mach}Exponentially fast diversity loss in the mother machine}

We extend our theory to quantify the rate of diversity loss in a mother machine. Conceptually, the mother machine is similar to our microchannels with one lane. The main difference is that, in the mother machine, reproductions can occur in only one direction since one end of the microchannel is sealed.  

We solve our model with one lane under such conditions, see Methods. In this case, we do not consider a mass effect, as reproduction event can only occur in one direction. We find that, for the mother machine, the diversity loss is still given by \eqref{A_t_one} and the average fixation time by  \eqref{fix_time_m}, where we set $m=0$ in both expressions. These results show that the change in boundary conditions do not affect the dynamics of diversity loss. 

\subsection*{\label{sec:Fixat_probab}Cells in the middle of a microchannel possess a positional advantage}

We expect cells located far from the open ends of the microchannel to benefit from a positional advantage. We quantify this idea by means of the fixation probability $P^{\fix}_{i,j}$, defined as the probability that the clonal population whose initial ancestor has coordinates $i,j$ eventually takes over the microchannel. In the one-lane case and for large $N$, the fixation probability is approximated by
\begin{equation}
    P^{\fix}_{i}=\phi_{m,N}(i) ,
\end{equation}
where $\phi_{m,N}(i)= \exp[-(i-\mu)^2/(2\sigma^2)]/\sqrt{2\pi\sigma^2}$ is a Gaussian distribution with mean $\mu = (N-1)/2$ and variance $\sigma^2 = (1-m/2)(N-1)/4$, see SI Appendix. This means that, at increasing the mass effect, mutants that are likely to take over the population are located in a narrower region at the center of the microchannel, see Fig.~S3G. In particular, the value of the mass parameter that we estimated ($m=0.6$) leads to a 30\% reduction in $\sigma^2$, compared with the case $m=0$. 

\begin{figure}[ht!]
\centering
\includegraphics[width=.90\linewidth]{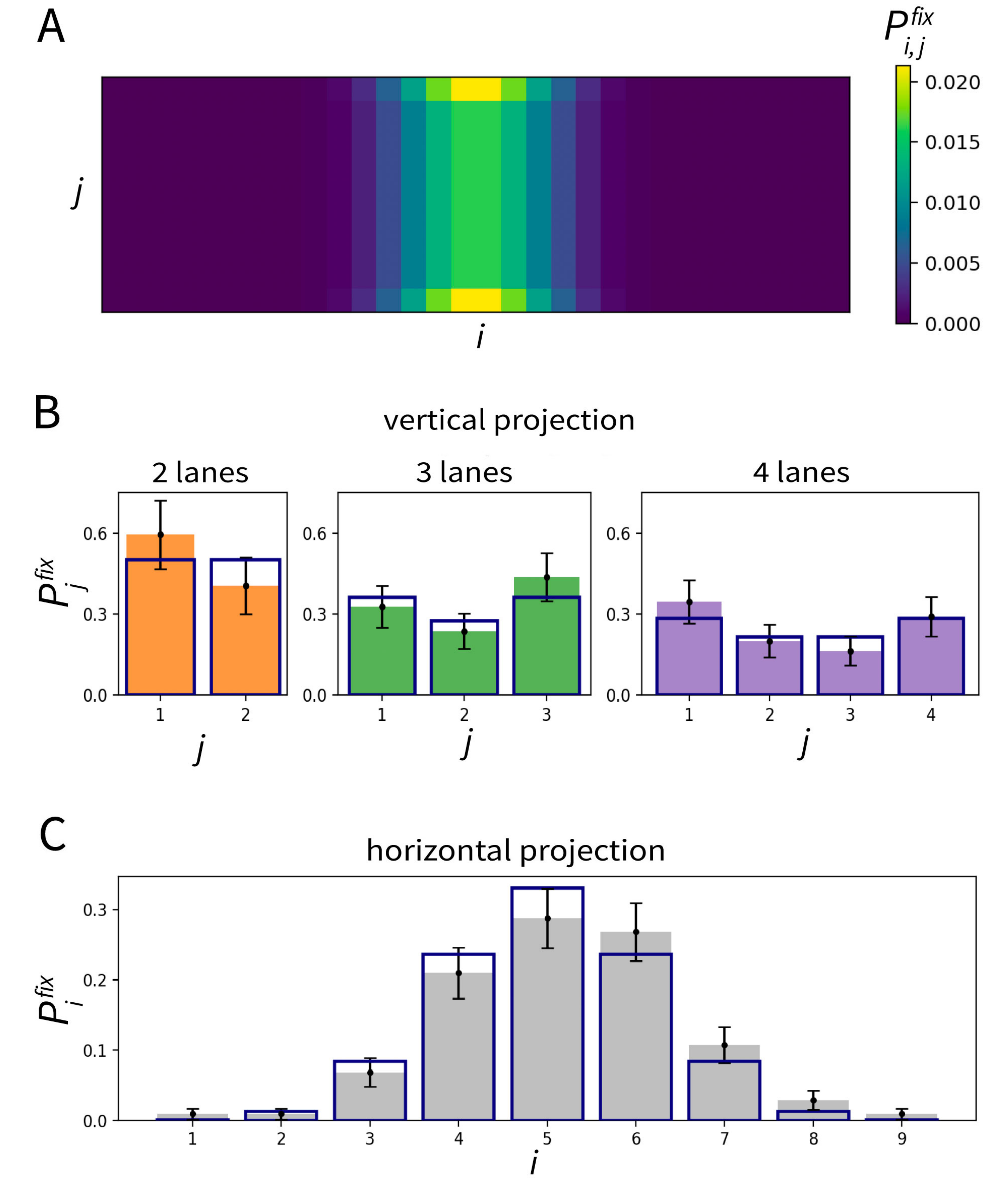}
\caption{Fixation probabilities are highest at the center of a microchannel. \textit{(A)} Fixation probabilities predicted by \eqref{P_ij} for $M = 10$, $N = 30$, $m = 0.6$ and $\alpha = 3.2$. \textit{(B)} Fixation probabilities along the vertical ($j$) axis.  Color histograms represent empirical probabilities from experiments with associated uncertainties. Dark blue bars represent marginalized fixation probabilities $P^{\fix}_j = \sum_i P^{\fix}_{i,j}$, where $P^{\fix}_{i,j}$ is given in  \eqref{P_ij}. The number of cells per lane is $N=9$ for all three cases. In the experiments where populations do not reach fixation, we use all remaining clonal populations at the end of the experiment (typically, from 2 to 6) to approximate the empirical fixation probabilities. The validity of this approximation is supported by numerical simulations, see SI Appendix. \textit{(C)} Projections of the fixation probabilities along the horizontal ($i$) axis. Grey histogram represents the empirical fixation probabilities. Dark blue bars represent $P^{\fix}_i = \sum_j P^{\fix}_{i,j}$.}
\label{fig:fix_prob}
\end{figure}

In microchannels with multiple lanes, we approximate the fixation probabilities  by
\begin{equation}
P^{\fix}_{i,j} \approx
    \begin{cases}
      \begin{aligned}
      &\frac{2\alpha+6}{M(2\alpha\!+\!3)\!+\!6}~ \phi_{m,N}(i) & \text{if}\ j=1, M,\\
      &\frac{2\alpha+3}{M(2\alpha\!+\!3)\!+\!6}~ \phi_{m,N}(i) & \text{otherwise},
    \end{aligned}
    \end{cases}
    \label{P_ij}
\end{equation}
see Fig.~\ref{fig:fix_prob}A. The approximation in \eqref{P_ij} is valid in the limit of large $N$ as well. We also require the two regimes of diversity loss to be well separated, see SI Appendix. The fixation probabilities predicted by \eqref{P_ij} are in good agreement with experimental observations, see Fig.~\ref{fig:fix_prob}B and \ref{fig:fix_prob}C.

\section*{Discussion}

In this paper, we studied the population genetics of microbial populations growing in microchannels with open ends. We base our study on a simple model in which cells are placed in lanes that are shifted by reproduction events. This shifting dynamics, combined with the geometry of the microchannels, causes a fast exponential loss of genetic diversity, rapid fixation within each lane, and slower competition among lanes.  Our experiments quantitatively confirm these predictions and reveal that population dynamics generates stripes of clonal populations inside the microchannels. This outcome is in sharp contrast to the case of bacterial populations growing on agar plates, where competing populations organize into sectors whose boundaries perform random walks \cite{hallatschek2007genetic}. 

Previous population models considered shifting dynamics in a single lane. Allen and Novak studied a similar model on a one dimensional ring \cite{allen2012evolutionary}, concluding that this dynamics does not significantly affect the selection strength. However, a model of epithelial tissues has led to the conclusion that the shift dynamics suppresses selection \cite{Iwasa}. In this latter model, cells are arranged in a linear array and can be expelled only on one end, in a similar fashion as in the mother machine \cite{rob_Ecoli}. Therefore, random mutations of the cell next to the opposite end are very likely to reach fixation, whereas fixation probabilities of mutations occurring elsewhere are very small.  Our results show that this imbalance in fixation probabilities is a robust consequence of pushing dynamics in channels with multiple lanes and open ends, that does not require a strict geometric constraint imposing all cells to descend from a mother cell.

Our results can be extended to populations growing in wider and deeper channels. In such populations, the ordered lane structure is disrupted at large scales, potentially leading to jamming \cite{delarue2016self,bi2016motility}. In fact, crowding is known to affect the outcome of competition among microbial strains even in non-confined geometries \cite{giometto2018physical,kayser2019collective}. Clarifying the connection between these evidences and our results is an interesting venue for future studies.

Our findings are potentially relevant for microbial ecology in soil. Bacteria in soil are spatially organized into relatively isolated and confined microenvironments \cite{o2007visualization}, where fluid flows supply cells with nutrients \cite{micr_compet_soil}. Most bacteria colonize micropores with diameter of about three times their body size \cite{kilbertus1980microhabitats} and in any case smaller than 6 $\mu$m, since small pores retain water for longer time. Moreover, bacteria residing in a pore of appropriate size are sheltered against larger predators \cite{young2000tillage}. Although bacteria have very large population sizes, it was estimated that each bacterium in soil interacts with about 120 other individuals on average  \cite{raynaud2014spatial}. Taken together, these observations support that the size of our microchannels is comparable to that of typical bacterial microenvironments in soil.

Renewing epithelial tissues in multicellular organisms \cite{marshak2001} present a similar spatial organization as the one studied in our work. Two main examples are the epidermis  \cite{janes2002epidermal} and the intestinal crypt \cite{bach2000stem}. The epidermis consist of multiple compartments of cells that originate from a stem cell layer.  These stem cells divide and generate differentiated cells that are shifted toward the top of the compartment, and can be eliminated from the tissue once they reach the surface. In the intestine, stem cells divide at the bottom of each crypt, move upwards, differentiate and are removed once they reach the top of the villus. This dynamics permits to rapidly expel cells that have accumulated deleterious mutations, thereby decreasing the risk of cancer which would otherwise have high chances to arise in rapidly growing epithelial tissues \cite{history_of_cancer,frank2007dynamics}. Similarly, proliferation of intestinal stem cells is thought to be disregulated in carcinogenesis \cite{bach2000stem}. Our findings can potentially be extended to understand evolutionary dynamics of these tissues.
In particular, adapting our model to study cancer dynamics would require introducing non-neutral clonal populations, as fitness differences between cell types is important for cancer progression  \cite{Outcompeting_cancer, Cell_Competition}.  Moreover, cells that carry tumour-promoting mutations can be eliminated by apoptosis, or killed by surrounding cells, so that more detailed cellular interactions should be considered in this case.

The fact that a constrained geometry has such a drastic impact on population genetics should be taken into account when designing experimental evolutionary studies. Our results open possibilities for constraining evolution by shaping the geometry of a microchannel hosting a microbial population.

\section*{Materials and Methods}
\subsection*{\label{subsec:exper_set}Bacteria strains and maintenance}
MG1655, a derivative of \textit{E. coli} K-12 wild-type strain, was used in this study. We transformed a plasmid in MG1655 to constitutively express green fluorescence protein (GFP) (pUA66 PrpsL-GFP Kan$^{R}$)  \cite{Rational_Design}. For the red fluorescent  strain, the pUA66 plasmid with mCherry open reading frame replacing that of GFP was customly constructed (VectorBuilder, USA). We cultured the MG1655 hosting pUA66 plasmids in Luria-Bertani (LB) broth (Lennox) supplemented with 50~$\mu$g mL$^{-1}$ Kanamycin. Detailed description of bacteria culture and plasmid engineering is provided in SI Appendix.

\subsection*{Microfluidic chip design and microfabrication protocols} The poly(dimethylsiloxane) (PDMS) device (on top of the cover glass) consisted of the top flow channels  for nutrients delivery and bacteria removal, and the bottom growth channels for bacteria growth, removal, and monitoring, see SI Appendix, Fig.~S2A. In order to perfuse fresh nutrients and create flow to remove bacteria, twenty growth channels of different width were intersected with 16 flow channels (L$\times$W$\times$H=4500$\times$50$\times$15$\mu$m). Growth channels (L$\times$H$=$30$\times$1$\mu$m) with varying width (1, 1.5, 2.5, and 3~$\mu$m) and interspacing of 10~$\mu$m were designed such that bacteria could be expelled from both ends, in contrast to the single-end design of the mother machine \cite{rob_Ecoli, monitoring}. The flow channels were joined by flow equalizing tree-like channels and flow resistors on both ends \cite{Design_microfluidic} and connected to an inlet and an outlet.

A silicon mold with the microstructures designed as above was fabricated by multi-step lithography with negative photoresist and maskless direct writing. PDMS microfluidic devices (Sylgard 184, Dow Corning, USA) were fabricated by standard soft lithography \cite{Soft_Lithography}. The PDMS slab was first cut and punched with an inlet and outlet using a puncher, then bonded to a high-precision No.1.5H cover glass using plasma activation. A 2~mm-thick acrylic frame was cut with CO$_2$ laser cutter  and affixed on top of the PDMS as a reservoir for bacteria seeding, completing the fabrication of the integrated microdevice. Detailed information of design and microfabrication is provided in SI Appendix.

\subsection*{Bacterial lineage tracking and time-lapse microscopy} Log-phase MG1655 \textit{E. coli} harboring plasmids for fluorescent proteins were grown at $37^{\circ}$C with vigorous shaking in LB broth supplemented with antibiotic kanamycin until the optical density at 600~nm (OD$_{600}$) reached 0.2. The bacteria suspension was concentrated 20 times by centrifugation before being injected into the PDMS microdevice which was pre-treated with a passivation solution to reduce bacterial binding to the microdevice surfaces. The PDMS microdevice was mounted in a microscope on-stage incubator pre-equilibrated at $37^{\circ}$C on an inverted motorized epi-fluorescence microscope. The bacteria were allowed to enter the growth channels under static condition for 2 hours before the M9 media were infused using a syringe pump first at a flow rate of 1.6~$\mu$L~min$^{-1}$ and doubled every two hours until it reached 16~$\mu$l~min$^{-1}$.

The fluorescence images were taken with either a 100X oil immersion objective or a 60X oil immersion objective with 1.5X intermediate magnification at an interval of $\Delta t=3$ minutes using high sensitivity camera (Prime95B, Photometrics, Canada) with GFP or mCherry filter cubes. More details are provided in SI Appendix. 

\subsection*{\label{subsec:data_process}Image analysis and data processing}
We processed the time-lapse recordings of our experiments using ImageJ software \cite{imagej}.
We used MicrobeJ plugin to detect bacteria in each frame \cite{microbeJ} and custom Python-program to track all the bacteria in time. The tracking algorithm is based on construction and comparison of local structures for each cell, see SI Appendix, Fig.~S3 A. We then reconstructed spatial lineage trees for each cell in the channel, see SI Appendix, Fig.~S3 B.

In experiments with multiple lanes,  microscopy focus drift issues for long-term live-cell imaging experiments caused occasional quality loss in the recordings for a few frames. We cropped the recordings when such issues occurred. The duration of our recording after this operation are  $13.4\pm 4.7$, $9.7\pm 2$, and $7.93\pm 1.5$, corresponding to $5.64 \pm 2.02$, $5.84 \pm 1.35$, $4.76 \pm 0.95$ generations for the experiments with $2, 3$ and $4$ lanes, respectively.  

To explore the second regime, we perform a re-tracking of the experimental data (see SI Appendix). In the re-tracking, we consider an initial condition in which each lane is occupied by a single clonal population. Re-tracking is justified by the observation that, during the time at which $A(t) = M$ in our original tracking, at least 70\% of each lane is occupied by a single clonal population, see SI Appendix.  

\subsection*{\label{subsec:the_model}Reproduction rates and model parameters}
We assign to each cell its coordinates $(i,j)$, with $1\le i\le N$ and $1\le j\le M$. In the following, we refer to the coordinate $i$ as the ``horizontal'' or ``axial'' coordinate.   We define the reproduction probability $k_{(i',j')(i,j)}$ as the probability that the daughter of a cell at position $i,j$ is placed at position $i',j'$.

\begin{figure*}[ht!]
\centering
\includegraphics[width=\textwidth]{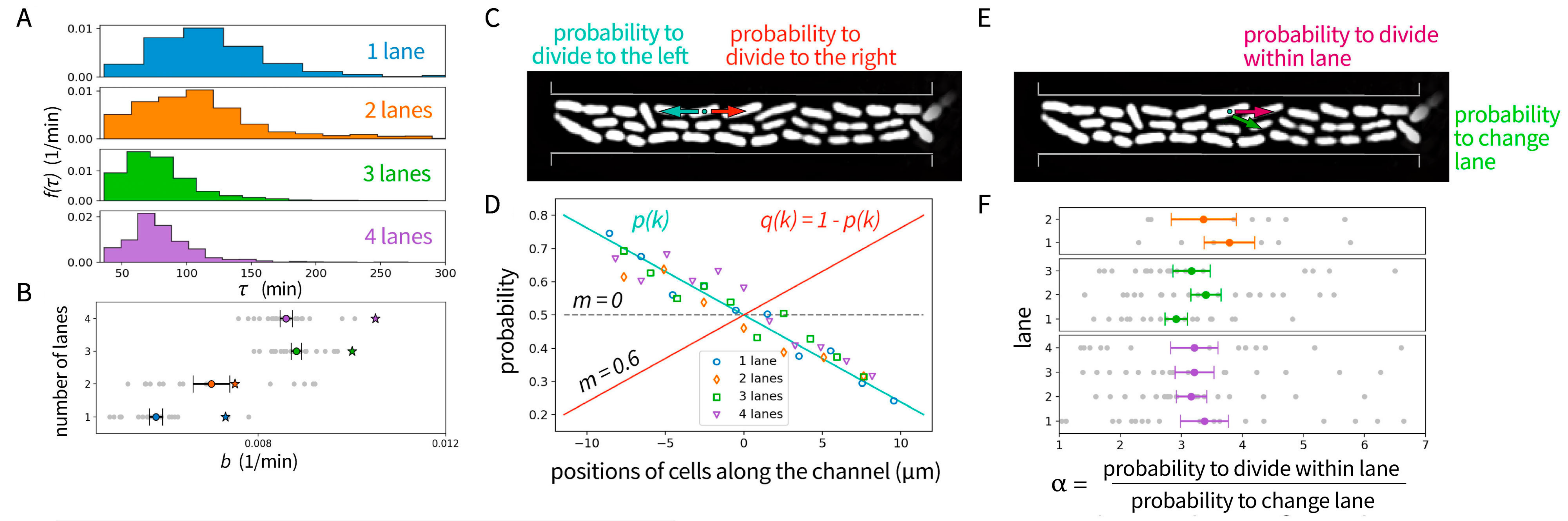}
\caption{Estimation of the model parameters from experimental observations. \textit{(A)} Empirical distributions of division times. \textit{(B)} Growth rates evaluated from the experimental data. The error bars represent the mean values and the standard errors of the population growth rates obtained as $\ln(2)/\langle \tau \rangle$, where $\langle \tau \rangle$ is the cell division time averaged over the population in a single experimental run. The stars mark the reproduction rates $b$ evaluated by solving the Euler--Lotka equation $2\langle \exp(-b\tau)\rangle=1$, see SI Appendix. \textit{(C)} Scheme of the two possible directions of division. We interpret an asymmetry between frequencies of reproductions in these two directions as a mass effect. \textit{(D)} Mass effect in the experimental data. The scattered points represent the average frequencies of leftward divisions as a function of the cell position in the experimental data . We fit the data with the linear function given in \eqref{mass_mod_1D} using the least squares method, resulting in $m = 0.6$ for all channel widths. \textit{(E)} Division within the same lane and to a neighbor lane. The alignment parameter $\alpha$ is defined as the ratio between the probability of a cell division within a lane to that of a cell division involving a change of lane. \textit{(F)} Average value and the standard error of $\alpha$ estimated from the experimental data. We find that $\alpha$ does not significantly vary across experiments and lanes. The average value over all experiments is $\alpha = 3.2$. The averages are calculated over 17 microchannels with 2 lanes, 21 microchannels with 3 lanes, and 20 microchannels with 4 lanes.  }
\label{fig:m_alpha}
\end{figure*}

We start from the case with one lane, $M=1$. In this case, the probabilities $p(i)=k_{i-1,i}$ and $q(i)=k_{i+1,i}$ that a cell at  position $i$ reproduce to the left and right, respectively, are expressed by
\begin{equation}
p(i) = -\!\frac{im}{N\!-\!1}+\frac{m(N\!+\!1)\!+\!N\!-\!1}{2(N\!-\!1)}, \quad q(i) = 1 \!- \!p(i) .
\label{mass_mod_1D}
\end{equation}
where $m$ is the mass parameter. These probabilities depend linearly on $i$, as observed in experiments (see Fig.~\ref{fig:m_alpha}D).

In the case $M>1$, we determine the reproduction probabilities by imposing two constraints. First, the probabilities to divide to the left $k_{(i-1,j)(i,j)}$ and to the right $k_{(i+1,j)(i,j)}$ satisfy the condition $k_{(i-1,j')(i,j)}/k_{(i+1,j')(i,j)} = p(i)/q(i)$ for $j'=(j-1,j,j+1)$. Second, the ratio between the probability of a cell division within a lane to that of a cell division involving a change of lane must be equal to $\alpha$:  $k_{(i',j)(i,j)}/k_{(i',j')(i,j)} = \alpha$ with $i'=(i-1,i+1)$ and $j'=(j-1,j+1)$.

The reproduction probabilities for cells in the bulk of the population ($1<j<M$) satisfying these constraints are expressed by
\begin{equation}
\begin{split}
k_{(i-1,j)(i,j)} &= \alpha \frac{-2mi+m(N+1)+N-1}{2(N-1)(\alpha+3)}, \\
k_{(i+1,j)(i,j)} &= \frac{2\alpha}{2(\alpha+3)} - k_{(i-1,j)(i,j)}, \\
k_{(i-1,j-1)(i,j)}  &= k_{(i-1,j+1)(i,j)}  = \frac{k_{(i-1,j)(i,j)} }{\alpha},\\
k_{(i+1,j-1)(i,j)} &= k_{(i+1,j+1)(i,j)} = \frac{k_{(i+1,j)(i,j)} }{\alpha},\\
k_{(i,j-1)(i,j)}  &= k_{(i, j+1)(i,j)}  = \frac{1}{2(\alpha+3)}.
\end{split}
\label{gen_mod_inner}
\end{equation}
If $i'=i$, the lane of cells $j'$ is shifted either to the left or right with probabilities given by \eqref{mass_mod_1D}.

We impose the same constrains for cells next to the top boundary of the microchannel ($j=1$), obtaining
\begin{equation}
\begin{split}
k_{(i-1, 1)(i,1)} &= \alpha \frac{-2mi+m(N+1)+N-1}{(N-1)(2\alpha+3)},\\
k_{(i+1, 1)(i,1)} &= \frac{2\alpha}{2\alpha+3}-k_{(i-1, 1)(i,1)}, \\
k_{(i-1, 2)(i,1)} &= \frac{k_{(i-1, 1)(i,1)}}{\alpha},\\
k_{(i+1, 2)(i,1)} &= \frac{k_{(i+1, 1)(i,1)}}{\alpha},\\
k_{(i,2)(i,1)} & = \frac{1}{2\alpha+3}. \\
\end{split}
\label{gen_mod_bound}
\end{equation}
The reproduction probabilities for cells next to the bottom boundary ($j = M$) can be similarly expressed. In all cases, if the $i$ coordinate of the daughter is equal to $0$ or $N+1$, she is immediately expelled from the microchannel. 

The model parameters evaluated from experiments are summarized in Table~\ref{table2}. For each microchannel width, we estimated $M$ and $N$ as the average number of lanes and the average number of cells per lane in our experiments, respectively. See SI Appendix for details. The estimation of parameters $b$, $m$, and $\alpha$ is detailed in Fig.~\ref{fig:m_alpha}.

\subsection*{\label{sec:loss_of_gen_divers}Dynamics of interfaces}

\begin{figure}[ht!]
\centering
\includegraphics[width=.75\linewidth]{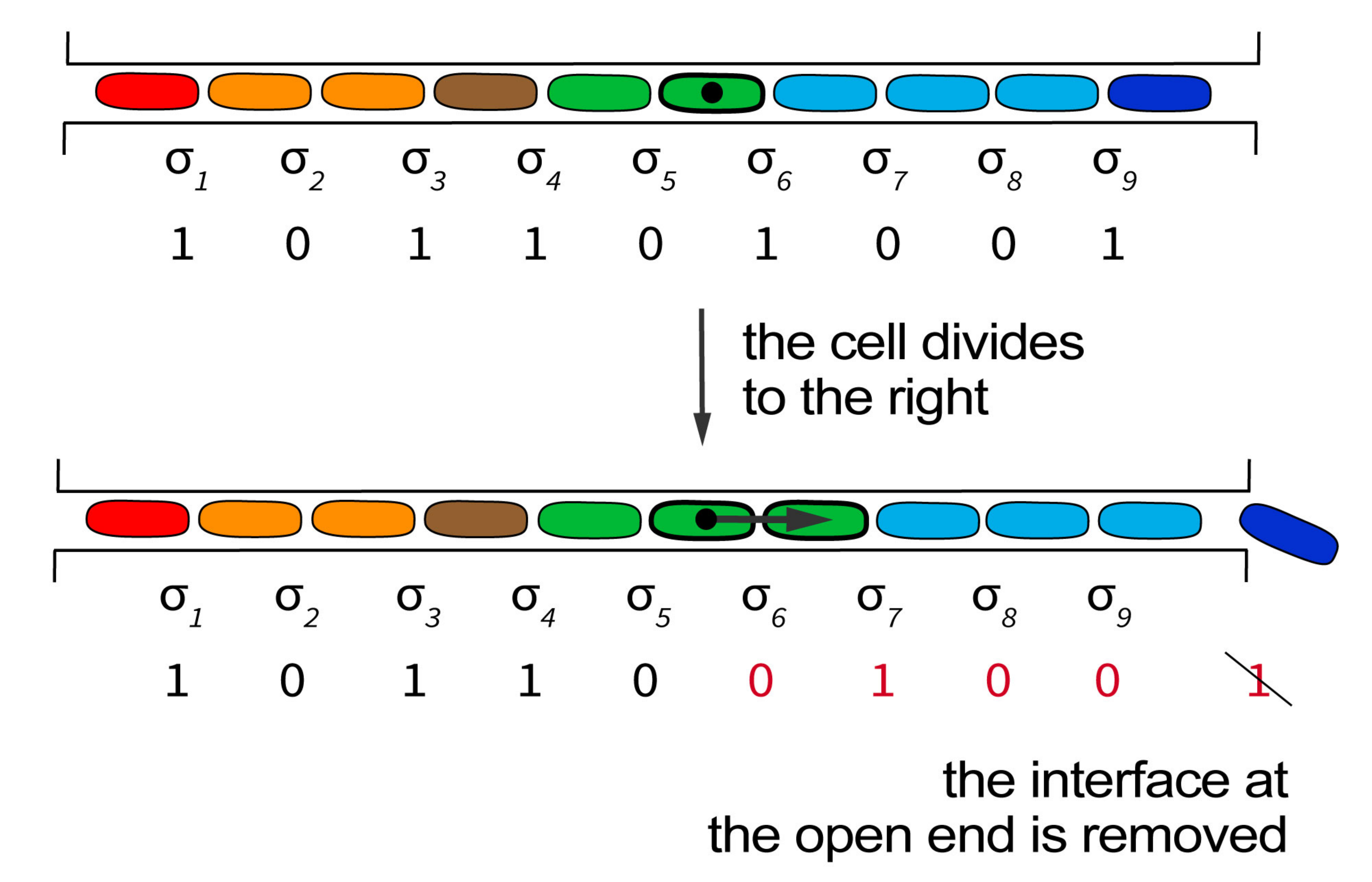}
\caption{Dynamics of interfaces. Different colors correspond to different clonal populations. Interfaces $\sigma_1, \sigma_2\dots \sigma_{N-1}$ are associated with adjacent cells. An interface $\sigma_i$ is equal to 1 if the two associated cells belong to different clonal populations and zero otherwise. As a consequence of cell division, an interface of value $0$ is created at $i=6$ and a portion of the vector of interface is shifted (in red). As an outcome, one interface is expelled.}
\label{fig:intefaces}
\end{figure}

We consider the case $M=1$ and assign to each cell at position $i$ at time $t$ the position $f_i(t)$ of its ancestor at time $t = 0$. The quantity $f_i(t)$ changes every time the cell at position $i$ is replaced by another one having a different initial ancestor. Two neighboring cells having the same value of $f_i(t)$ are conspecific, i.e. they belong to the same clonal population. We assign interfaces to neighboring cells that are not conspecific. We encode these interfaces into a vector $\vec{\sigma}(t) = (\sigma_1(t), \sigma_2(t), \dots, \sigma_{N-1}(t))$, whose components are defined by
\begin{equation}
\sigma_i(t) =
    \begin{cases}
      0 & \text{if}\ f_{i}(t) = f_{i+1}(t),\\
      1 & \text{if}\ f_{i}(t) \neq f_{i+1}(t).
    \end{cases}
    \label{sigma_bound}
  \end{equation}
The initial condition is $\vec{\sigma}(0) = (1, 1, \dots, 1)$. The vector of interfaces evolves until it reaches the absorbing state $(0, 0, \dots, 0)$ that corresponds to fixation of one clonal population. Each cell division creates a pair of conspecific cells and shifts all cells by one position, either to their right or left. This event implies that one interface $\{\sigma_i\}_{i = 1,\dots,N-1}$ is set to $0$ and all interfaces on one side of it are shifted by one position, depending on the direction of the cell division. As a consequence, an interface located at the open end may be removed from the vector, see Fig.~\ref{fig:intefaces}.

To describe the interface dynamics, we introduce inverse shift operators that take into account the presence or absence of an interface at the open end:
\begin{align}
\hat{b}^i_r \vec{\sigma} &= (\sigma_1, \sigma_2, \dots, \sigma_{i-1}, \sigma_{i+1}, \dots, \sigma_{N-1},0),\nonumber\\
\hat{b}^i_l \vec{\sigma} &= (0,\sigma_1, \sigma_2, \dots, \sigma_{i-1}, \sigma_{i+1}, \dots, \sigma_{N-1}),\nonumber\\
\hat{c}^i_r \vec{\sigma} &= (\sigma_1, \sigma_2, \dots, \sigma_{i-1}, \sigma_{i+1}, \dots, \sigma_{N-1},1),\nonumber\\
\hat{c}^i_l \vec{\sigma} &= (1, \sigma_1, \sigma_2, \dots, \sigma_{i-1}, \sigma_{i+1}, \dots, \sigma_{N-1}).
\label{shift_oper}
\end{align}
A state $\hat{b}^i_r \vec{\sigma}$ or $\hat{c}^i_r \vec{\sigma}$ evolves to a state $\vec{\sigma}$ if the $i$-th cell divides to the right. Similarly, a state $\hat{b}^i_l \vec{\sigma}$ or $\hat{c}^i_l \vec{\sigma}$ evolves to a state $\vec{\sigma}$ if the $(i+1)$th cell divides to the left. The master equation for the interface distribution is
\begin{align}
&\frac{dP_{\vec{\sigma}}}{dt} = -b\sum_{i=1}^{N-1}[q(i)+p(i+1)]P_{\vec{\sigma}}(t)+\nonumber\\
&+b\sum_{i=1}^{N-1}\delta_{\sigma_i,0}[q(i)(P_{\hat{b}^i_r \vec{\sigma}}+P_{\hat{c}^i_r \vec{\sigma}})+p(i+1)(P_{\hat{b}^i_l \vec{\sigma}}+P_{\hat{c}^i_l \vec{\sigma}})],
\label{mast_eqv}
\end{align}
where $p(i)$ and $q(i)$ are defined in \eqref{mass_mod_1D} and the Kronecker delta takes care of the fact that a reproduction event necessarily creates an interface of value equal to $0$. The solution of \eqref{mast_eqv} reads
\begin{equation}
P_{\vec{\sigma}} = \prod_{i = 1}^{N-1}\left[\sigma_i e^{-\beta t}+(1-\sigma_i)(1-e^{-\beta t})\right],
\label{mast_solut}
\end{equation}
where $\beta$ is defined in \eqref{beta}. The solution given in \eqref{mast_solut} can be verified by direct substitution into \eqref{mast_eqv}. 
\eqref{mast_solut} shows that the interfaces $\{\sigma_i(t)\}_{i = 1, \dots, N-1}$ are independent, identically distributed random variables with $P(\sigma_i(t) = 1)  = e^{-\beta t}$ for all $i$. 
The diversity is related to the number of interfaces by
\begin{equation}
A(t) = \sum_{i=1}^{N-1}\sigma_i(t)+1.
\label{eq:at_interf}
\end{equation}
Computing the average of $A(t)$ using \eqref{mast_solut} leads to
\eqref{A_t_one}.

\subsection*{\label{sec:fixat_time_calcul}Fixation time in the model with one lane}

We calculate the fixation time for the model with $M = 1$, employing the interface formalism. Removing one interface amounts to removing one clonal population.  We denote by $T_{A\rightarrow A-1}$ the time it takes to remove the $A$th clonal population. The fact that the interfaces $\sigma_i(t)$ are independent, identically distributed random variables implies that the time intervals $T_{A\rightarrow A-1}$ are exponentially distributed with mean $\langle T_{A\rightarrow A-1}\rangle=1/[\beta(A-1)]$. This observation directly implies \eqref{fix_time_m}.

\subsection*{\label{sec:fixat_time_calcul_MM}Fixation time in the mother machine}

We apply the interface formalism to a model of the mother machine. Without loss of generality, we assume that the open end is the right one. Therefore, the vector of interfaces evolves according to a master equation including only right inverse shift operators:
\begin{equation}
\frac{dP_{\vec{\sigma}}}{dt} = -b(N-1)P_{\vec{\sigma}}(t)+b\sum_{i=1}^{N-1}\delta_{\sigma_i,0}(P_{\hat{b}^i_r \vec{\sigma}}+P_{\hat{c}^i_r \vec{\sigma}}),
\label{mast_eqv_mm}
\end{equation}
see \eqref{shift_oper} and \eqref{mast_eqv}. The solution to this master equation also factorizes:
\begin{equation}
P_{\vec{\sigma}} = \prod_{i = 1}^{N-1}[\sigma_i e^{-b t}+(1-\sigma_i)(1-e^{-b t})].
\label{mast_solut_mm}
\end{equation}
It follows that \eqref{fix_time_m} with $\beta=b$ holds for the mother machine as well.

\begin{acknowledgements}
We gratefully acknowledge support of the Okinawa Institute of Science and Technology Graduate University (OIST) with subsidy funding from the Cabinet Office, Government of Japan. We are grateful for the help and support provided by the Scientific Computing
and Data Analysis section of Research Support Division at OIST. We  thank colleagues from the Micro/Bio/Nanofluidics Unit at OIST: Daniel Carlson for assistance in numerical simulation and Riccardo Funari for fruitful discussion.  We thank Professor Paola Laurino from OIST for gifting the MG1655 strain, and Dr. Pamela Silver from Harvard University for gifting the pUA66 PrpsL-GFP plasmid.  We thank Deepak Bhat, Massimo Cencini, Kirill Korolev, and Robert Ross for feedback on a preliminary version of the manuscript.
\end{acknowledgements}

\bibliography{zHenriquesLab-Mendeley}

\onecolumn
\newpage

\section*{Supplementary Information}

\section*{Experimental setup and characterizations}
\subsection*{Bacteria strains and maintenance}\hfill

In this study, we used MG1655, an \textit{E.~coli}  strain derived from K-12 wild-type strain. To allow single bacteria tracking in microchannels, a low-copy plasmid harboring the promoter sequence for 30S ribosomal protein and kanamycin resistance (Kan$^{\mbox{R}}$) cassette was used to constitutively express green fluorescent protein (pUA66 PrpsL-GFP Kan$^{\mbox{R}}$) (Addgene plasmid \#105606; \url{http://n2t.net/addgene:105606}; RRID: Addgene\_105606) \cite{Rational_Design}.
To visualize both green and red fluorescent markers simultaneously, a pUA66 plasmid harboring rpsL promoter with mCherry open reading frame was constructed by VectorBuilder (vector ID:VB200629-1195vze, vectorbuilder.jp). The plasmid was transformed in MG1655 using standard chemical transformation with ice-cold 0.1 M calcium chloride and 42$^{\circ}$C heat shock \cite{Preparation}. For long term cryopreservation, we stored log phase bacteria suspension supplemented with glycerol to a final concentration of 20~wt\% in $-80^{\circ}$C.

\subsection*{Microfluidic chip design and microfabrication} \hfill

{\em Microfluidic chip design aided by numerical simulation.} The 3D microchannel design was created by AutoCAD (Autodesk, USA) and exported into COMSOL Multiphysics (COMSOL Inc, USA) with tetrahedral mesh for numerical simulation to guide the design before the microfabrication step, see Fig.~\ref{fig:S2}A. The creeping flow module was used and 10~mm$^3$ min$^{-1}$ flow rate was set to simulate the culture media flowing in the microfluidic channel. The flow rate distribution in the mid-plane of growth channels was used to optimize the flow equalizing channel design (Fig.~\ref{fig:S2} B). To further characterize the flow at the openings of the growth channel, the flow velocity and shear rate at mid plane ($z = 0.5$~$\mu$m) of the 3~mm-wide growth channels were analyzed (Fig.~\ref{fig:S2}C). The flow velocity at $\pm$ 1 $\mu$m at the openings was $0.28 \pm 0.01$~mm s$^{-1}$ with shear rate of $691.3 \pm 23.8$~s$^{-1}$. The flow velocity in the growth channels, 5~$\mu$m from the openings was 5.1$\times$10$^{-4}$ $\pm$ 2.1$\times$10$^{-5}$~mm s$^{-1}$, with shear rate of  $0.29 \pm 0.01$~s$^{-1}$ (Fig.~\ref{fig:S2}D and E). These simulation results confirmed the high shear rate regime at the channel openings to enable the removal of old bacteria while keeping bacteria maintenance in the middle of the growth channels.

{\em Silicon mold microfabrication.}
Due to a difficulty in aligning microstructures embedded in the microdevice, a three-step fabrication procedure was employed. First, 5~$\mu$m-high negative photoresist (mr-DWL-5, Micro Resist Technology GmbH, Germany) was spin-coated on a piece of 4-inch silicon wafer (E\&M, Japan) using a spin coater (MS-B150, Mikasa, Japan). An alignment pattern containing cross-shaped microstructures was exposed at 405~nm using a maskless writer (DL-1000, Nanosystems Solutions, Japan). The structures were developed in propylene glycol methyl ether acetate (PGMEA, Sigma-Aldrich, USA) and washed thoroughly with isopropanol and distilled water before dried under nitrogen air. For the 1 $\mu$m-high growth channel layer, the mr-DWL-5 photoresist was diluted with gamma butyrolactone (B103608, Sigma-Aldrich, USA) at 7.5:1 (w/w) ratio and spin-coated on the silicon wafer containing alignment microstructures at 7500~rpm for 30~s. After appropriate soft bake based on manufacturer’s protocol (95$^\circ$C for 5 ~min), the patterns for the growth layer channels were exposed using the DL-1000 maskless writer with 2X subpixel exposure (170 mJ~cm$^{-2}$) followed by a post exposure bake at 95$^{\circ}$C for 5~min, finally cooled down to room temperature.

Without development of the growth channel layer, the mr-DWL-5 photoresist for the flow channel layer (15~$\mu$m) was spin-coated on the post exposure-baked mold and soft-baked according to manufacturer’s protocol. The flow channel pattern was aligned and exposed using the same maskless writer. After post-exposure hard bake at 95$^{\circ}$C for 5~min, the mold was developed in PGMEA solution and washed thoroughly with isopropanol and distilled water before dried under nitrogen air. The true heights of the microstructures were confirmed with a stylus profilometer (Dektak, Bruker, USA).

{\em Fabrication of PDMS microdevices.}
The patterned silicon mold was passivated with Trichloro (1H, 1H, 2H, 2H-perfluorooctyl) silane (Sigma-Aldrich, USA) fume in a vacuum desiccator for 2~h. The passivated mold was placed in a custom mold made in-house with polytetrafluoroethylene for casting 4~mm-high poly(dimethylsiloxane) (PDMS) silicone rubber on the mold. PDMS (Sylgard 184, Dow Corning, USA) was mixed and degassed at 10:1 = elastomer:curing agent ratio (Thinky ARE-310, Japan). Appropriate amount of PDMS was poured on the mold and cast with a cover of transparent acrylic sheet at 60$^{\circ}$C for at least 4~h. After PDMS crosslinking, individual piece was diced from the PDMS and 21~G holes were punched at the inlet and outlet using a puncher (Accu-punch MP, Syneo, USA). High-precision cover glasses (No.1.5H, 60$\times$24$\times$0.175 mm, Paul Marienfeld GmbH, Germany) were mounted on a 3D-printed washing stand made in-house and ultrasonically washed in 1\% TFD4 (Franklab, France) and distilled water before drying in an 80$^{\circ}$C oven. Two PDMS pieces were then bonded to a piece of cover glass using plasma surface activation (PDC-001-HP, Harrick Plasma, USA) at about 10$^{\circ}$ tilted angle to the long edge of the cover glass to avoid interference by microstructures to the optical focus locking mechanism of the microscope. After bonding, the two PDMS pieces on the cover glass were split using a diamond scriber (Ted Pella, USA). A piece of 2 mm-thick acrylic frame was cut with a CO$_2$ laser cutter (VLS2.30, Universal Laser Systems, USA) and adhered on the PDMS piece as a reservoir using a dual energy double sided tape (5302A, Nitto, Japan), completing the construction of the PDMS microdevice.

\subsection*{Chip culture and time-lapse microscopy} \hfill

{\em Preparation of microfluidic chips and media.}
To avoid microbubbles, the microfluidic channel was first passivated with 99.5\% ethanol (Nacalai Tesque, Japan), and replaced with ultrapure water (MilliQ, Millipore, USA) \cite{Voltage_gated}. A passivation solution composed of 2~mg mL$^{-1}$ bovine serum albumin (Nacalai Tesque, Japan) and 0.5~mg mL$^{-1}$ salmon sperm DNA (Thermo Fisher Scientific, USA) was incubated in the channel \cite{monitoring}.  A set of tubing with stainless tube (New England Small Tube, USA) connected to a 2.5~mL syringe (Terumo, Japan) was inserted in the outlet of the microfluidic chip. The passivation solution was infused in the microchannels by withdrawal with a syringe pump (YSP-202, YMC co., Ltd, Japan) at 2~$\mu$L min$^{-1}$ for 30~min.
To prepare the media for bacterial lineage experiments, 5X M9 salt stock (33.9~g L$^{-1}$ Na$_2$HPO$_4$; 15~g L$^{-1}$ KH$_2$PO$_4$; 2.5~g~L$^{-1}$ NaCl; 5~g L$^{-1}$ NH$_4$Cl) was autoclaved and diluted to 1X with supplementation of filter-sterilized MgSO$_4$, CaCl$_2$, glucose, kanamycin, rifapentine to final concentrations of 2~mM, 0.1~mM, 10~mM, 0.1~mM, 1~$\mu$M, respectively. The 1X M9 working solution was further supplemented with 100X MEM vitamins (M6895, Sigma-Aldrich, USA) at 1:100 dilution (1X) to formulate the M9 media. All chemicals were purchased from Nacalai Tesque Inc. Japan unless specified. The M9 media were loaded into two 25~mL glass syringes (SGE, Australia) and connected to a set of tubing with stainless tubes.

{\em Seeding bacteria on chip and time-lapse microscopy.}
MG1655 bacteria outgrown overnight in LB broth was inoculated to a new tube of 4~mL LB broth with 0.1~mM kanamycin. The \textit{E. coli} suspension grew at 37$^{\circ}$C with 200~rpm shaking until the suspension reached log phase with OD$_{600}$ of 0.2.  Bacteria suspension of 1~mL was centrifuged at 3000 $\times$~g for 10~min and resuspended in a 50~$\mu$L LB broth.
The microfluidic chip was affixed on a holder in an on-stage incubator (WKSM, Tokai Hit, Japan) on an inverted motorized epi-fluorescence microscope (Ti-E, Nikon, Japan) (see Fig.~\ref{fig:S2}F). A dummy microfluidic chip with a K-type thermocouple (ANBE MST Co., Japan) was affixed on the holder for temperature feedback and control.
The concentrated \textit{E. coli} suspension was injected in the microchannels using a micropipette and appropriate amount of passivation solution in the reservoir was added to equalize the hydrostatic pressure difference between the inlet and the outlet \cite{Voltage_gated}. The bacteria were allowed to swim into the growth channel and to grow for 2~h at 37$^{\circ}$C before the tubing of M9 media infusion was inserted into the inlet and the perfusion was started at 1.6~$\mu$L min$^{-1}$ using a multichannel syringe pump (neMESYS, Cetoni GmbH, Germany). The perfusion rate was doubled every 2~h until it reached 16~$\mu$L min$^{-1}$. The microchannels populated by cells were imaged in time-lapse microscopy. For visualization of bacterial population using MG1655 with GFP or mCherry fluorescent proteins, the imaging began after the initial incubation.
To start time-lapse microscopy, either an 100X oil immersion objective or a 60X oil immersion objective (Plan apo lambda, Nikon, Japan) with 1.5X intermediate magnification was used to take time-lapse images of the bacteria in the growth channels with focus locking assistance using the Perfect Focus System on a Nikon microscope. The bacteria were tracked using epi-fluorescence imaging using mercury lamp excitation (Intensilight, Nikon, Japan) with GFP or mCherry filter cubes (Semrock, USA). The images were taken on a scientific complementary metal oxide semiconductor camera (sCMOS, Prime95B, Photometrics, Canada) at an interval of 3~min and exported from the NIS Element software (Nikon, Japan). XY drift due to thermal noise and repeatability of the motorized stage was corrected when necessary using the Stackreg plugin in Fiji ImageJ \cite{imagej, stack_reg}.

\subsection*{Characterization of bulk bacteria growth}\hfill

In our experiments, we used \textit{E. coli} transformed with PrpsL-GFP or PrpsL-mCherry aiming to visualize the dynamics of two different strains. To validate that there was limited fitness difference between the two strains, we performed growth kinetics measurements for \textit{E. coli} strains with the two plasmids in LB media as well as the M9 media with 10~mM glucose (Fig.~\ref{fig:S2} G).
The MG1655 \textit{E.~coli} strains with the pUA66 plasmids were cultured in 4~mL LB broth with 0.1~mM kanamycin at 37$^{\circ}$C shaken in 200~rpm overnight. One $\mu$L of the suspensions were inoculated into fresh 1.5~mL LB broth or M9 medium with 10~mM glucose and 1X MEM vitamin, both supplemented with 0.1~mM kanamycin. The suspensions were placed in a 1.6~mL polystyrene cuvette (Sarstedt, Germany) and measured at OD$_{600}$ every minute in a UV spectrometer (UV-1800, Shimadzu, Japan). A temperature control unit was used to keep the cuvette at 37$^{\circ}$C. The measurements were performed in triplicates.

The growth kinetics of the two strains in LB and M9 media were similar. The best fit for both media was performed from 150~min  to 200~min. The growth rate constants $k$ in LB were 0.032~min$^{-1}$ for PrpsL-GFP and 0.029~min$^{-1}$ for PrpsL-mCherry, corresponding to a doubling time of $21.5 \pm  0.005$~min and $24.9 \pm  0.003$~min, which were similar to literature findings \cite{liang1999decay}.
The growth rate constants in M9 media were 0.0120~min$^{-1}$ for PrpsL-GFP and 0.0124~min$^{-1}$ for PrpsL-mCherry, corresponding to a doubling rate of 57.6~min and 55.8~min. The standard deviation of the residuals for fitting the growth curve of both strain was less than $4\,10^{-5}$.

\subsection*{\label{subsec:data_process_SI}Image analysis and data processing} \hfill

{\em Preprocessing.} We analyzed the time-lapse recordings of our experiments. The spatial resolution of each image was 0.11~micron/pixel with field of view of 1200$\times$1200 pixels. We filtered microchannels containing smoothly moving and proliferating bacteria during the entire recording using ImageJ software and its plugins \cite{imagej}.
We prepared the data for the processing in three steps. First, we rotated all images to horizontally align the boundaries of all microchannels. This step allows us to compare bacteria orientations among different recordings. Second, we corrected the recording drift using the plugin StackReg \cite{stack_reg}. Third, we applied the filter ``Unsharp Mask'' to reduce noise and increase image contrast.
From the preprocessed recordings, we determined contours of each bacteria in each frame.
We used MicrobeJ plugin that provides an automatic algorithm for detecting bacteria contours and centers using specified information about their morphology \cite{microbeJ}.

{\em Tracking.} The tracking algorithm identified the same bacteria between consequent frames.  Since we worked with densely packed bacteria, we used a tracking algorithm based on the construction of local structures and their comparison in consecutive frames \cite{tracking}. We constructed the local structures as Delaunay meshes with nodes at the centers of the bacteria (Fig.~\ref{fig:S3}A). The Delaunay mesh maximizes the minimum angle of the triangles that generate the mesh and it is independent on the topology of bacteria. In the algorithm, we compared cells using their positions with respect to other cells in the population rather than the morphological characteristics. Such approach is convenient for dense populations, because cells are in contact between each other. The algorithm encompasses a forward and a backward step:
\begin{itemize}
\item In the forward step, we associated a cell on a frame at time $t$  with the one on the next frame at time $t+\Delta t$  by minimizing the difference between
their local structures. 
\item In the backward step, all bacteria from a frame at time $t+\Delta t$ that were not associated with any bacteria on $t$ frame were compared to them in the same way in order to determine their mothers. 
\end{itemize}
Our algorithm is an adaptation of the algorithm originally proposed in Ref.~\cite{tracking}. Mainly, we modified the similarity measures between local structures and set the specific order of cells comparisons. For growing populations in microchannels with open ends, it was more beneficial to compare cells starting from the center of the microchannel toward open ends, because otherwise we risked making wrong associations at the beginning with the removed cell that could affect further performance of the algorithm, see Fig.~\ref{fig:S3}A. The tracking algorithm returned spatial lineage trees for each cell in the channel. An example of such tree is shown in Fig.~\ref{fig:S3}B. The spatial trees contain all the information about the population dynamics.

\section*{\label{sec:parameters}Estimation of model parameters}
\subsection*{\label{subsec:Populat_size}Number of lanes and number of cells per lane}\hfill

Our microchannels have the same length but different widths, thereby harboring from $1$ to $4$ different number of lanes of bacteria, see supplementary movie 4 and Table 1 in the Main Text. The number of cells within each lane fluctuates among different microchannels. We evaluate the average number of cells per lane as the total number of cells observed in each frame divided by the number of lanes. To calculate the average number of cells we analyzed 23 microchannels with 1 lane, 17 microchannels with 2 lanes, 21 microchannels with 3 lanes and 20 microchannels with 4 lanes.  For microchannels harboring $1, 2, 3$ and $4$ lanes, the number of cells per lane are $13 \pm 1$, $8.8 \pm 0.5$, $9 \pm 1$ and $9.2 \pm 0.5$, respectively. Microchannels with 1 lane harbor more cells because the width of the microchannels allows them to be slightly tilted and better packed within the same length (see supplementary movie 4). Accordingly, for our model, we fixed the four sets of parameters $(M, N) = (1, 13),~ (2, 9),~ (3, 9)$ and $(4, 9)$ for channels of width 1 $\mu$m, 1.5 $\mu$m, 2.5 $\mu$m, and 3 $\mu$m, respectively.

\subsection*{Division rate}\hfill

In our model, cells reproduce in age-independent manner with a constant division rate $b$, implying an exponential distribution of times between consecutive cell division.  However, the distribution $f(\tau)$ of division times of bacterial cells measured in experiments is markedly non-exponential. Here we define an effective division rate from such non-exponential distributions. In the model, since the division rate is constant, the average number of cells, including those that are expelled, grows exponentially as $n(t) = n(0)e^{b t}$. In the experiments, the total number of cells in our experiments grows exponentially as well, $n(t)=n(0) e^{\Lambda t}$, where $\Lambda$ is the population growth rate. The population growth rate is linked with the distribution $f(\tau)$ by the Euler--Lotka equation \cite{charlesworth2000fisher}
\begin{equation}
2\int_0^\infty  f(\tau) e^{-\Lambda \tau}d\tau= 1.
\label{euler_lotka}
\end{equation}
We determine the  division rate $b$ by imposing that the population growth rate should be the same in the model and in the experiments. This condition amounts to solve \eqref{euler_lotka} with $\Lambda=b$. We estimate the empirical distributions of the generation times $f(\tau)$ for each microchannel width from our experiments. We then estimate $\Lambda$ and thereby $b$ by numerically solving \eqref{euler_lotka} in each case. We find that the estimated division rate increases with the microchannel width. 
This effect is likely to be caused by stress  that limits bacterial growth rate in narrower microchannels. A positive correlation between channel width and growth rate was previously observed in single-lane mother machines \cite{yang2018analysis}, consistent with our results. The authors of Ref.~\cite{yang2018analysis} interpreted this effect as a mechanical impediment to growth.

We also verified that, as assumed in the model, the average reproduction rate does not significantly change with the position of cells along the microchannels (see Fig.~\ref{fig:S3} C) . 

The division rates evaluated with the Euler--Lotka equation are significantly larger than the estimate $b=\Lambda=\ln 2/\langle \tau\rangle$, that one obtains in the case of constant division times.  The reason is that, in the presence of fluctuations, fast-reproducing cells provide a larger contribution to the population growth rate than slow-reproducing ones \cite{hashimoto}.

\subsection*{\label{subsec:weak_inter}Test of the existence of two regimes of diversity loss}\hfill

According to our assumptions, the first regime of diversity loss is characterized by competition within each lane. We then expect, at the end of the first regime ($A(t) = M$), that each lane should be approximately dominated by one species. We experimentally tested this assumption and found that, during the time at with $A(t)=M$, at fraction ranging  71--95~\% of each lane is occupied by one strain, see Table~\ref{tab:lanes}. This result support the validity of our approximation and justifies the retracking procedure we used to explore the dynamics of the second regime of diversity loss.

We run numerical simulations of the model with $\alpha = 3.2$, $m = 0.6$, $N = 10$, and $M = 2, 3, 4$. We observe formation of stripes of dominant strains in the model as well, see Table \ref{tab:lanes}.

\begin{table}[h]
\centering
\caption{\label{tab:lanes}Average percentage $\pm$ SE of a dominant strain within each lane during the time at which $A(t) = M$ measured from experiments}
\begin{tabular}{c|c|c|c|c}\\
number of lanes&1st lane&2nd lane&3rd lane&4th lane\\ \hline
 $M = 2$&$0.95 \pm 0.02$&$0.98 \pm 0.02$&$-$&$-$\\
  $M = 3$&$0.89 \pm 0.05$&$0.7 \pm 0.04$&$0.76 \pm 0.05$&$-$\\
$M = 4$&$0.86 \pm 0.04$&$0.8 \pm 0.04$&$0.71 \pm 0.05$&$0.85 \pm 0.04$
\end{tabular}
\end{table}

\section*{Mathematical derivations}

\subsection*{\label{subsec:first_regime}First regime of diversity loss} \hfill

Our first step is to compute the distribution of the fixation time $T_{N \to 1}$ for $M=1$. The fixation time can be expressed by $T_{N \to 1} = \sum_{A=2}^{N} T_{A\rightarrow A-1}$, where the time it takes to expel the $A$th clonal population $T_{A\rightarrow A-1}$ is exponentially distributed with mean $\langle T_{A\rightarrow A-1}\rangle=1/[\beta(A-1)]$, where $\beta = b\left(1-\frac{m}{N-1}\right)$.
Therefore, $T_{N \to 1}$ is hyperexponentially distributed:
\begin{equation}
f(T_{N \to 1})=\sum_{n=1}^{N-1} \prod_{j = 1; j\ne n}^{N-1}\frac{j}{j-n}ne^{-\beta n T_{N \to 1} }
=\sum_{n=1}^{N-1}(-1)^{n-1}{\binom{N}{n}}n e^{-\beta n T_{N \to 1}}
=(N-1)(1-e^{-\beta T_{N \to 1}})^{N-2}e^{-\beta T_{N \to 1}}.
\label{dens_fn}
\end{equation}
The cumulative distribution function reads
\begin{equation}
F(T_{N \to 1}) = \int_0^{T_{N \to 1}} f(x)dx = (1-e^{-\beta T_{N \to 1}})^{N-1}.
\label{distr_fn}
\end{equation}

We now move to the case $M>1$. If the parameter $\alpha$ is large enough, lanes evolve independently, each of them eventually reaching its fixation state. We calculate the average time at which all $M$ lanes have reached their fixation states (first regime). We call $T_{N \to 1}^j$ the time at which lane $j$ has reached fixation. In terms of these times, the duration of the first regime of diversity loss is equal to $T_{MN \to M}= \max(T_{N \to 1}^1, T_{N \to 1}^2, \dots, T_{N \to 1}^M)$.

In this case, \eqref{distr_fn} governs the distribution of the fixation time $T_{N \to 1}^j$ of each lane independently.
Therefore, the distribution of the time $T_{MN \to M}$ at which all lanes have reach fixation is expressed by
\begin{equation}
g(T_{MN \to M}) = M f(T_{MN \to M})F^{M-1}(T_{MN \to M})= M(N-1)e^{-\beta T_{MN \to M}  }(1-e^{-\beta T_{MN \to M}  })^{M(N-1)-1}.
\end{equation}
The average of $T_{MN \to M}$ is expressed by
\begin{equation}
\langle T_{MN \to M} \rangle = \int_0^\infty zg(z)dz
=M(N-1)\int_0^\infty z e^{-\beta z}(1-e^{-\beta z})^{M(N-1)-1} dz
= \beta^{-1} \sum_{k = 1}^{M(N-1)} \frac{1}{k} \approx  \beta^{-1} [\log(M(N-1)) + \gamma],
\label{X_aver}
\end{equation}
where $\gamma$ is the Euler--Mascheroni constant. 

\subsection*{\label{subsec:second_regime}Second regime and fixation time in the case with multiple lanes}\hfill

Once each lane has reached fixation, a second regime initiates in which lanes compete with each other. We describe this competition dynamics by a linear invasion process, see Fig.~\ref{fig:S3} D. We call $p_{j}$ the rate at which a clonal population in a lane $j$ successfully invades a neighboring lane.  This rate is equal to the sum of the individual rates at which each  cell in the lane can invade a neighboring lane and then reach fixation. We express these individual rates, in turn, as products of rates for a cell to reproduce in a neighboring lane times the fixation probabilities in the new lane. Following this logic we obtain:
\begin{equation}
p_{j} =
    \begin{cases}
      \begin{aligned}
      &\frac{3(N-1)-4m}{(N-1)(2\alpha+3)}b & \text{if}\ j = 1, M,\\
      &\frac{3(N-1)-4m}{2(N-1)(\alpha+3)}b & \text{otherwise}.
    \end{aligned}
    \end{cases}
    \label{weights}
  \end{equation}

In deriving \eqref{weights}, we have used that the fixation probabilities within one lane are normalized and therefore their contribution disappears when summing over all cells in a lane. The characteristic time $1/p_j$ should be interpreted as the typical interval between successful invasion events, but does not include the time required for the invading population to fixate in the new lane.

We now calculate the distributions of the time intervals $T_{A \to A-1}$, for $A = M, \dots, 2$  using ~\eqref{weights}. These intervals terminate when a cell from a lane successfully invade another lane and then reaches fixation in the new lane. We already know that the distribution of the fixation time in one lane  with $N$ cells is given by a sum of exponentially distributed random variables with mean $1/[\beta(k-1)]$, $k = N, \dots, 2$. In the case of fixation within one lane in the multiple-lane model, we multiply the rates $1/[\beta(k-1)]$ by the rate of reproduction of cells within one lane that are equal to $2\alpha/(2\alpha+3)$ for boundary lanes, and $\alpha/(\alpha+3)$ for inner lanes.

Therefore, the distribution of the time interval $T_{A \to A-1}$ for the model with $M$ lanes is given by a sum of $N$ exponentially distributed random variables. We call $\lambda^{A, M}_1, \lambda^{A, M}_2, \dots, \lambda^{A, M}_{N}$ the $N$ rates characterizing these random variables. One random variable, with rate $\lambda^{A, M}_1$,  is associated with a successful invasion. The remaining $(N-1)$ random variables  have rates $\lambda^{A, M}_2, \dots, \lambda^{A, M}_{N}$ and are associated with fixation in the invaded lane. Thus, we express the density function of $T_{A \to A-1}$ as
\begin{equation}
f_{T_{A \to A-1}}(x) = \Big[ \prod_{i = 1}^N \lambda^{A, M}_i \Big] \sum_{j = 1}^N \frac{e^{-x\lambda^{A, M}_j }}{\prod\limits_{\substack{k = 1 \\ k \ne j}}^{N} (\lambda^{A, M}_k - \lambda^{A, M}_j)}.
\label{dens_hyperep}
\end{equation}
Therefore we obtain
\begin{equation}
P(A(t) = M) = P(t < T_{M \!\to \!M-1}) \!=\!1-\!\int_0^t \!f_{T_{M \!\to\! M-1}}(x) dx\! =1 \!- \!\Big[ \prod_{i = 1}^N \lambda^{M, M}_i \Big]\! \sum_{j = 1}^N \!\frac{1 - e^{-t\lambda^{M, M}_j }}{\lambda^{M, M}_j\!\prod\limits_{\substack{k = 1 \\ k \ne j}}^{N} (\lambda^{M, M}_k \!- \!\lambda^{M, M}_j)}.
  \label{P_At_is_M}
\end{equation}
For $A = M$, the rates are
\begin{equation}
\lambda^{M, M}_1 =
    \begin{cases}
      \begin{aligned}
      &2\frac{3(N-1)-4m}{(N-1)(2\alpha+3)} & \text{if}\ M = 2,\\[10pt]
      &\frac{(4\alpha+9)(3(N-1)-4m)}{(N-1)(\alpha+3)(2\alpha+3)} & \text{if}\ M = 3,\\[10pt]
      &2\frac{(3\alpha+6)(3(N-1)-4m)}{(N-1)(\alpha+3)(2\alpha+3)} & \text{if}\ M = 4,
    \end{aligned}
    \end{cases}
    \label{lam_1_Mmn1}
  \end{equation}

\begin{equation}
\lambda^{M, M}_k =
    \begin{cases}
      \begin{aligned}
      &\frac{2\alpha(k-1)}{2\alpha+3}\Big(1-\frac{m}{N-1}\Big) & \text{if}\ M = 2,\\[10pt]
      &\frac{4\alpha(k-1)}{4\alpha+9}\Big(1-\frac{m}{N-1}\Big) & \text{if}\ M = 3,\\[10pt]
      &\frac{3\alpha(2\alpha+5)(k-1)}{2(3\alpha+6)(\alpha+3)}\Big(1\!-\!\frac{m}{N-1}\Big) & \text{if}\ M = 4,
    \end{aligned}
    \end{cases}
    \label{lam_k_Mmn1}
  \end{equation}
for $k = 2, \!\dots, \!N$.

For  $A = M-1$, the rates have the form
\begin{align}\label{lam_1_Mmn2}
\lambda^{2, 3}_1 &= \frac{3(N-1)-4m}{2(N-1)(\alpha+3)} \\
\lambda^{3, 4}_1 &=\frac{(16\alpha^2+63(\alpha+1))(3(N-1)-4m)}{6(N-1)(\alpha+3)(\alpha+2)(2\alpha+3)} \nonumber
\end{align}
and
\begin{align} \label{lam_k_Mmn2}
\lambda^{2, 3}_k &=\frac{2\alpha(k-1)}{2\alpha+3}\Big(1-\frac{m}{N-1}\Big) \\
\lambda^{3, 4}_k &=(k\!-\!1)\left[\frac{\alpha(4\alpha\!+\!9)\!(6\alpha\!+\!\!15)}{4(3\alpha\!+\!6)^{\!2}\!(\alpha\!+\!3)}\!\!+\!\!\frac{\alpha}{3\alpha\!\!+\!\!6} \right]\Big(1\!\!-\!\!\frac{m}{N\!\!-\!\!1}\Big), \nonumber
\end{align}
for $k = 2, \!\dots, \!N$.
Thus, we have
\begin{equation}
 P(A(t) = M-1)  = P(T_{M \to M\!-\!1} \leq t,  T_{M \to M\!-\!1} + T_{M\!-\!2 \to M\!-\!2} > t)   = 
 \int_{0}^t \int_{t}^{\infty} f_{T_{M \to M\!-\!1}, T_{M \to M\!-\!1} + T_{M\!-\!1 \to M\!-\!2}} (y_1, y_2) dy_1 dy_2.
\end{equation}
We now express the joint density function by
\begin{equation}
  \begin{split}
f_{T_{M \to M\!-\!1}, T_{M \to M\!-\!1} + T_{M\!-\!1 \to M\!-\!2}} (y_1, y_2) &= f_{T_{M\! \to\! M\!-\!1}} (y_1) f_{T_{M\!-\!1 \to M\!-\!2}} (y_2 \!- \!y_1) =\\
& = \prod_{i = 1}^N \lambda^{M, M}_i \prod_{i = 1}^N \lambda^{M\!-\!1, M}_i \! \sum_{j = 1}^N\! \frac{e^{-\lambda^{M, M}_j y_1}}{\prod\limits_{\substack{k = 1 \\ k \ne j}}^{N} (\lambda^{M, M}_k\!-\!\lambda^{M, M}_j)}\! \sum_{j = 1}^N \frac{e^{-(\lambda_j^{M-1, M})^2 (y_2\!- \!y_1)}}{\prod\limits_{\substack{k = 1 \\ k \ne j}}^{N} (\lambda^{M\!-\!1, M}_k \!-\! \lambda^{M\!-\!1, M}_j)},
\end{split}
\label{P_At_is_Mmn1}
\end{equation}
where $\{\lambda^{M, M}_i\}_{i = 1,\dots,N}$ are given by \eqref{lam_1_Mmn1}, \eqref{lam_k_Mmn1}, and $\{\lambda^{M-1, M}_i\}_{i = 1,\dots,N}$ are given by \eqref{lam_1_Mmn2}, \eqref{lam_k_Mmn2}.

The distributions of the time interval $T_{M-2 \to M-3}$ is hyperexponential  with  density given by \eqref{dens_hyperep} as well. In this case, the rates are given by
\begin{equation}
  \begin{split}
&\lambda^{M-2, 4}_1 = \frac{(3(N-1)-4m)}{2(N-1)(\alpha+3)}, \\
&\lambda^{M-2, 4}_k =\frac{2\alpha(k-1)}{2\alpha+3}\Big(1-\frac{m}{N-1}\Big), \quad  k = 2, \dots, N.
\label{lam_k_Mmn3}
\end{split}
\end{equation}

\subsection*{\label{subsec:fixat_time}Fixation time in the case $M>1$}\hfill

We now estimate the fixation time for $M = 2, 3, 4$ lanes as a sum of two contributions. The first contributor follows from the first regime. This is the time at which each lane reaches its fixation state, given by \eqref{X_aver}. The second contribution comes from competition between the lanes, with an average duration that can be calculated from its representation as a sum of independent exponentially distributed random variables with the rates given by \eqref{lam_1_Mmn1}--\eqref{lam_k_Mmn2} and \eqref{lam_k_Mmn3}.

For $M = 2$, the fixation time is expressed as a sum of these three contributions by
\begin{equation}
\langle T_{2N\to 1}\rangle \approx  \frac{1}{\beta}[\log(2(N\!-\!1)) + \gamma] + \frac{1}{b}\sum_{k = 1}^N \frac{1}{\lambda^{M-1}_k}.
\label{fix_tm_2ln}
\end{equation}
For $M = 3$, 
\begin{equation}
\langle T_{3N\to 1}\rangle \approx  \frac{1}{\beta}[\log(3(N\!-\!1)) + \gamma] + \frac{1}{b}\sum_{k = 1}^N \frac{1}{\lambda^{M-1}_k} + \frac{1}{b}\sum_{k = 1}^N \frac{1}{\lambda^{M-2}_k}.
\label{fix_tm_3ln}
\end{equation}
For $M = 4$, 
\begin{equation}
\langle T_{4N\to 1}\rangle \approx  \frac{1}{\beta}[\log(4(N\!-\!1)) + \gamma]+\frac{1}{b}\sum_{k = 1}^N \frac{1}{\lambda^{M-1}_k} + \frac{1}{b}\sum_{k = 1}^N \frac{1}{\lambda^{M-2}_k} + \frac{1}{b}\sum_{k = 1}^N \frac{1}{\lambda^{M-3}_k}.
\label{fix_tm_4ln}
\end{equation}

The dependence of the fixation time on the parameter $\alpha$ is represented in Fig.~\ref{fig:S3} E.

\subsection*{\label{subsec:fixat_probab}Fixation probabilities}

In this section, we prove that the fixation probability for the cell at the $i$th position in the case $M=1$ has the form 
\begin{equation}
P_i^{\fix} =
    \begin{cases}
      \begin{aligned}
      &\frac{{\binom{N-1}{i-1}}}{2^{N-1}} & \text{if}\ m = 0,\\
      &\frac{\binom{N-1}{i-1} \binom{2(N-1)/m-N-1}{(N-1)/m-i-2}}{\binom{2(N-1)/m-2}{(N-1)/m-1}}\ & \text{if}\ m \neq 0,
      \end{aligned}
    \end{cases}
    \label{P_k}
\end{equation}
with $i = 1, \dots, N$. This result directly leads to Eq.~[7] in the Main Text, upon approximating the binomial distributions appearing in Eq.~(\ref{P_k}) with Gaussian distributions in the limit of large $N$.

A clonal population originating at the $i$th position reaches its fixation state if and only if $(i-1)$ cells on its left are expelled from the left end and $(N-i)$ cells on its right are expelled from the right end. Therefore, to find the fixation probability, we calculate the probabilities that the two groups of cells at the left and right of the chosen cell are expelled.  We introduce a Markov process $n(t)$ which represents the number of cells to the left of the chosen clonal population, see Fig.~\ref{fig:S3} F.  This process has two absorbing states: $n(t) = 0$ implies that the chosen clonal population has reached the left end of the microchannel;
$n(t) = N$ implies that the chosen clonal population went extinct.
We write the master equation for the process $n(t)$ as
\begin{equation}
    \frac{dP_n(t)}{dt} = r_{n+1 \rightarrow n} P_{n+1}(t)+  r_{n-1 \rightarrow n}P_{n-1}(t)
    - (r_{n \rightarrow n+1}+r_{n \rightarrow n-1})P_n(t), \label{mast_eq_1D}
\end{equation}
where the transition rates are defined by
\begin{equation}
\begin{split}
r_{n \rightarrow n+1} &= b\sum_{i = 1}^n  q(i) = b\frac{n[N-1-m(N-n)]}{2(N-1)},\\
r_{n \rightarrow n-1} &= b\sum_{i = 1}^n p(i) = b\frac{(N-n)[N-1 -mn ]}{2(N-1)}.
\end{split}
\label{rates_nt}
\end{equation}
We rewrite \eqref{mast_eq_1D} with the initial condition $n(0) = n_0$, $n_0 = 1, \dots, N$ in the matrix form
\begin{equation}
    \begin{cases}
      \begin{aligned}
      &\partial_t \boldsymbol{P}(t) = M \boldsymbol{P}(t), \\
      &\boldsymbol{P}(0) =\boldsymbol{P}_0, \\
    \end{aligned}
    \end{cases}
    \label{sys_diff}
  \end{equation}
where $\boldsymbol{P}(t) = (P_1(t), P_2(t), \dots, P_N(t))^T$. The initial condition is $P_n(t=0)=\delta_{n,n_0}$. The matrix  $M$ is a tridiagonal transition matrix
with the following elements on the upper diagonal
\begin{equation}
M_{n,n-1}=b_n = -\frac{b}{2}(N-n-1)\Big(1-\frac{m(n+1)}{N-1}\Big),
\end{equation}
on the main diagonal
\begin{equation}
M_{n,n}=a_n = \frac{b}{2}\Big(N-2mn\frac{N-n}{N-1}\Big),
\end{equation}
and on the lower diagonal
\begin{equation}
M_{n,n+1}=c_n = -\frac{b}{2}n\Big(1-m\frac{N-n}{N-1}\Big).
\end{equation}
The propagator for the process $n(t)$  is expressed by
\begin{equation}
P_{n|n_0}(t) = (e^{Mt}\boldsymbol )_{n,n_0}.
\label{solut_elem}
\end{equation}
Therefore, the probability flux to the absorbing state $N$ at time $t$ has the form
\begin{equation}
J_{N|n_0}(t) = r_{N-1 \rightarrow N}P_{N-1|n_0}(t) = \frac{b(N-1-m)}{2}(e^{Mt})_{N-1,n_0}.
\end{equation}
Thus, the probability of being absorbed in $N$ given the initial state $n_0 = 1, \dots, N$ is equal to
\begin{equation}
P_{N|n_0} \!\! = \!\!\!\int_{0}^{\infty}\!\!\!\!J_{N|n_0}(t) dt \! =\!-\frac{b(N\!-\!1\!-\!m)}{2}(M^{-\!1})_{N\!-\!1,n_0}.
\label{P_Nl}
\end{equation}
To write this probability in an explicit form, we use a result for the inverse of a tridiagonal matrix \cite{usmani}. The $(N-1, k)$th element of the inverse of the tridiagonal transition matrix $M$ is given by
\begin{equation}
(M^{-\!1})_{N\!-\!1,k} = -\frac{2}{b}
    \begin{cases}
      \begin{aligned}
     &(-\!1)^{N\!-\!1\!+\!k}  \frac{\phi_N \theta_{k-1}}{\theta_{N-1}} \prod_{i=k}^{N-2}\!c_i  \mbox{  for }  k\!<\!N\!-\!1, \\
      &\theta_{N-2}\frac{\phi_N}{\theta_{N-1}} \mbox{ for }  k = N-1,\\
    \end{aligned}
        \end{cases}
\end{equation}
where
\begin{equation}
\theta_k = a_k\theta_{k-1}-b_{k-1}c_{k-1}\theta_{k-2}, \quad i = 2,3,\dots,N-1,
\end{equation}
\begin{equation}
\phi_k = a_k\phi_{k+1}-b_{k}c_{k}\phi_{k+2}, \quad i = N-2, N-3,\dots,1,
\end{equation}
with the boundary values $\theta_0 = 1$, $\theta_1 = a_1$, $\phi_N = 1$, $\phi_{N-1} = a_{N-1}$. Since the upper and lower diagonals are symmetric, $\theta_k = \phi_{N-k}$. Using these expressions, we obtain
\begin{equation}
\phi_k = \frac{(N-1)!}{(k-1)!}\prod_{i=1}^{N-k}u_i+(N-k)!\sum_{i=k}^{N-1}\binom{N-1 }{ i} \prod_{j=0}^{N-i-1}u_j\prod_{j=k}^{i}u_j,
\label{phi_mass}
\end{equation}
where $u_i = (1-mi/(N-1))$.
\eqref{P_Nl} and ~\eqref{phi_mass} imply that
\begin{equation}
P_{N|n_0} = \frac{\sum_{i=0}^{n_0-1}\binom{N-1 }{ i}\prod_{j=0}^{N-i-1}u_j\prod_{j=0}^iu_j}{\sum_{i=0}^{N-1}\binom{N-1}{ i}\prod_{j=0}^{N-i-1}u_j\prod_{j=0}^iu_j}.
\label{P_N_k}
\end{equation}
An analogous result holds for the group of cells to the right of the chosen clonal population, due to symmetry. Using these absorbing probabilities, we write the fixation probability as
\begin{equation}
P_i^{\fix} =P_{N|i} - P_{N|i-1}.
\label{P_k_form}
\end{equation}
After some algebra, using ~\eqref{P_N_k} and \eqref{P_k_form} we obtain \eqref{P_k}. The theoretical result \eqref{P_k} is confirmed by numerical simulations, see Fig.~\ref{fig:S3} G.

\subsection*{\label{subsec:emp_fixat_probab}Empirical fixation probabilities}\hfill

We estimate the empirical fixation probabilities from all the clonal populations that are left in the microchannels at the end of the experiments (typically, from 2 to 6).  To verify that this approximation does not significantly affect our estimate, we numerically simulate the model for the time as the averaged duration of the experimental recordings in each case: $5.6, 5.8$ and $4.7$ generations for $M = 2, 3, 4$, respectively. As a result, we observe $2 - 7$ competing populations in most of the cases. The resulting fixation probabilities are very close to the theoretical ones given by Eq. (7) in the Main Text, see Fig.~\ref{fig:S3} H, I.



 \begin{figure}
 \centering
 \includegraphics[width=.75\linewidth]{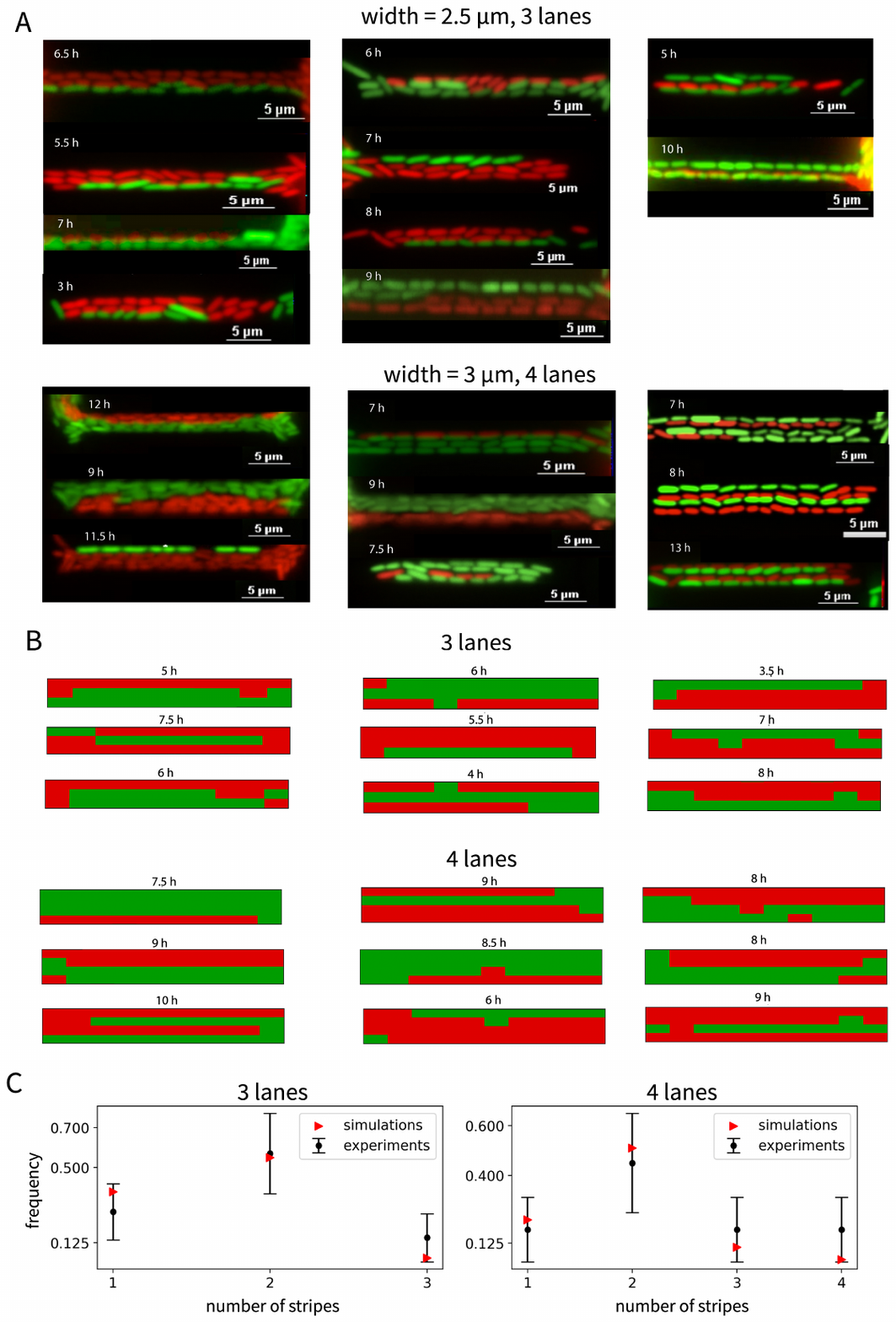}
 \caption{\textit{(A)} Experimental patterns formed by two neutral \textit{E.coli} strains (in red and green). In microchannels of width 2.5~$\mu$m, among 14 experimental realizations, we observe 2 cases with 3 stripes, 8 cases with 2 stripes and 4 cases with a single stripe (corresponding to fixation, not shown). In microchannels of width 3~$\mu$m, among 11 experimental realizations, we observe 2 cases with 4 stripes, 2 cases with 3 stripes and 5 cases with 2 stripes and 2 cases with 1 stripe (not shown). The snapshots are taken at the times when the pattern is formed. The average times are $7.2 \pm 2.1$ hours and $8.7 \pm 2.9$ hours for the microchannels of width 2.5~$\mu$m and 3~$\mu$m, respectively. \textit{(B)} Snapshots of numerical simulations of the model with two strains and  $M=3, 4$ . Other parameters are the same as in Fig.~2B in the Main Text. At the initial time $t = 0$, the population is constituted by two randomly mixed strains with equal frequencies. The snapshots are taken at the times  when the pattern starts forming, i.e. when there are dominant strains (with frequency $\geq$ 70\% within the lane) in each lane. The average of these times are
$6.5 \pm 3.2$ hours and $8.3 \pm 3.4$ hours  for $M=3$ and 4, respectively.  \textit{(C)} Distribution of the number of stripes. Black error bars represent the experimental observations shown in  \textit{(A)}. Red triangles represent the numerical simulations as in panel \textit{(B)}. In this case, we ran $5000$ simulations for each case ($M=3,4$).  }
 \label{fig:S1}
 \end{figure}

\begin{figure}
\centering
\includegraphics[width=.90\linewidth]{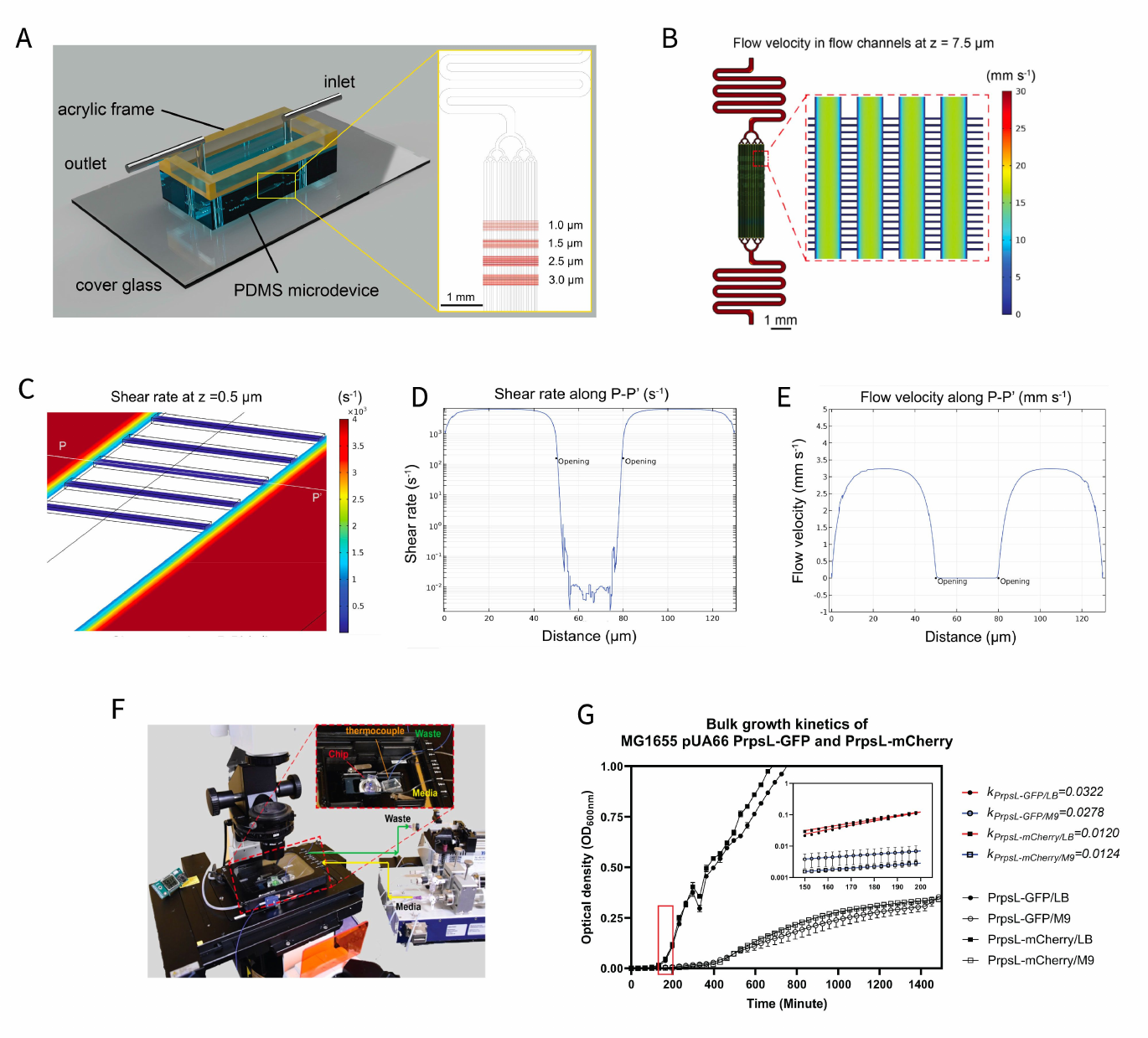}
\caption{\textit{(A)} Composition of the integrated microdevice used to track bacteria lineage in real time. The inset shows half of the symmetric design of microstructures in the PDMS device. The red pattern indicates the growth channels and the black pattern indicates the flow channels that are joined by a tree-like flow stabilizing structure and a flow resistor to an inlet or outlet. \textit{(B)} Numerical simulation of the fluid flow in our microdevice. The dashed inset in red shows the magnified region in the chip where the flow channels have high flow velocity ($16.2 \pm 0.3$~mm s$^{-1}$). \textit{(C)} Shear rate at the mid-plane (z = 0.5 $\mu$m) of growth channel, as obtained from numerical simulation of fluid flow at the intersection of flow channels and growth channels. \textit{(D, E)}  Data from the transverse line along P and P’ on the side wall of the flow channels were extracted for quantification. \textit{(F)} Photo image of the microscopy setup. The inset shows the PDMS microdevice placed in the on-stage temperature incubator. The M9 media was infused using a multichannel syringe pump and the waste was withdrawn using a syringe pump at early stage before the flow rate reached 16 $\mu$l min$^{-1}$ and then left open to atmosphere. \textit{(G)} Growth kinetics of MG1655 with pUA66 PrpsL-GFP plasmid or pUA66 PrpsL-mCherry plasmid in bulk LB broth or M9 media were monitored by OD$_{600}$ at every minute. To improve visualization, the plot is downsampled to one point every 30 minutes. The inset shows the early log phase between 150~minutes and 200~minutes. The growth curves were fitted with an exponential function $Y=Y_0 \exp(kt)$. The growth rate constants $k$ in LB were 0.0322$\pm$0.0012~min$^{-1}$ for PrpsL-GFP and 0.0278$\pm$0.0008~min$^{-1}$ for PrpsL-mCherry, corresponding to a doubling time of 21.5~minutes and 24.9~minutes, respectively ($P=0.2$, Student's \textit{t} test). The growth rate constants in M9 media were 0.0120$\pm$0.0001~min$^{-1}$ and 0.0124$\pm$0.0004~min$^{-1}$ for PrpsL-GFP and PrpsL-mCherry, corresponding to doubling time of 57.6 and 55.8~minutes ($P=0.29$, Student's \textit{t} test).}
\label{fig:S2}
\end{figure}

\begin{figure}
\centering
\includegraphics[width=\textwidth]{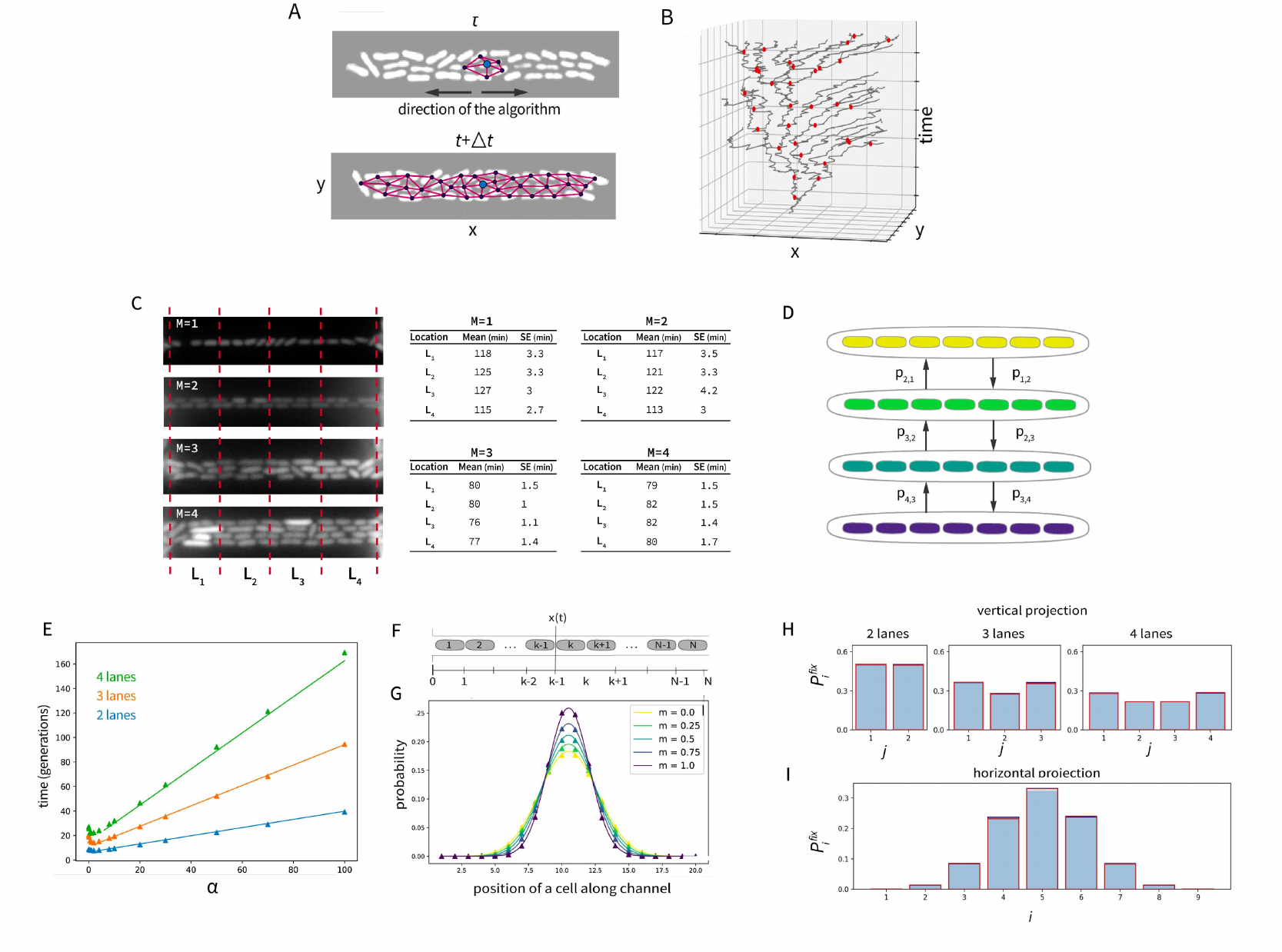}
\caption{\textit{(A)} Local structures of cells constructed on two consequent images from time-lapse experimental recordings of proliferating \textit{E. coli} in a microchannel. The local structure of a cell from the top image is compared to the local structures of cells from the bottom image. \textit{(B)} An example of a spatial generation tree of a clonal population constructed from the experimental data after the tracking step. Grey trajectory represents positions of clonal cells in time, red points represent division events. \textit{(C)} Average division time of bacteria depending on their position along the microchannels. \textit{(D)} Sketch of  competition among lanes in the second regime of diversity loss. Lanes are shown in different colors. Arrows represent successful invasion rates, given by ~\eqref{weights}. \textit{(E)} Fixation time as a function of the parameter $\alpha$. Solid lines represent the theoretical solutions given in ~\eqref{fix_tm_2ln}, \eqref{fix_tm_3ln}, \eqref{fix_tm_4ln} for $b = 1$ and $m = 0$. Triangle-shape points represent numerical simulations for $m = 0$, $M = 2, 3$ and $4$. \textit{(F)} Scheme of the process $n(t)$ given by ~\eqref{rates_nt}. At each time point, $n(t)\in[0, \dots, N]$. If $n(t)$ reaches the absorbing state $0$ or $N$, the population at the right (left) end fixates.  \textit{(G)}  Fixation probabilities for the model with one lane of $N = 20$ cells. Solid curves represent fixation probabilities given in Eq.~\eqref{P_k}, the triangle-shape points represent numerical simulations of the model with $N = 20$ and different values of $m$. \textit{(H)} Fixation probabilities along the vertical ($j$) axis.  The histograms show results of numerical simulations of the model with the parameters $\alpha = 3.2$, $m = 0.6$, $N = 9$ and $M = 2, 3, 4$. We run the simulations for $5.6, 5.8$ and $4.7$ generations for $M = 2, 3, 4$, respectively. Dark red bars represent results of numerical simulations  of the model with the same set of parameters, that we run until populations reach fixation. Dark blue bars represent marginalized theoretical fixation probabilities $P^{\fix}_j = \sum_i P^{\fix}_{i,j}$. 
\textit{(I)} Projections of the fixation probabilities along the horizontal ($i$) axis. The histograms are the result of numerical simulations with the same parameters as in \textit{(G)}. Dark blue bars represent $P^{\fix}_i = \sum_j P^{\fix}_{i,j}$.}
\label{fig:S3}
\end{figure}


\end{document}